\documentclass[superscriptaddress,reprint,amsmath,amssymb,aps,floatfix]{revtex4-1}

\usepackage{amsmath,gensymb,textcomp,bm,dcolumn,eurosym,array,tabu,multirow,nicefrac,color,subfigure,graphicx,upgreek}
\usepackage[colorlinks, linkcolor=blue, citecolor=blue, urlcolor=blue, breaklinks=true]{hyperref}

\usepackage{mathpazo,times} 

\newcommand{\ket}[1]{\lvert #1 \rangle}

\usepackage{bbm}
\usepackage[subsectionbib]{bibunits}
\usepackage{graphicx}
\usepackage{dcolumn}
\usepackage{bm}
\usepackage{units}
\usepackage[british]{babel}
\usepackage{psfrag}
\usepackage{float}
\usepackage{color}

\begin{document}


\title{Temporal distinguishability in Hong-Ou-Mandel interference: Generation and characterization of high-dimensional frequency entanglement}

\author{Yuanyuan Chen}
\affiliation{
Institute for Quantum Optics and Quantum Information - Vienna (IQOQI), Austrian Academy of Sciences, Boltzmanngasse 3, 1090 Vienna, Austria.}
\affiliation{Vienna Center for Quantum Science \& Technology (VCQ), Faculty of Physics, University of Vienna, Boltzmanngasse 5, 1090 Vienna, Austria}
\affiliation{Department of Physics and Collaborative Innovation Center for Optoelectronic Semiconductors and Efficient Devices, Xiamen University, Xiamen 361005, China}

\author{Sebastian Ecker}
\affiliation{
Institute for Quantum Optics and Quantum Information - Vienna (IQOQI), Austrian Academy of Sciences, Boltzmanngasse 3, 1090 Vienna, Austria.}
\affiliation{Vienna Center for Quantum Science \& Technology (VCQ), Faculty of Physics, University of Vienna, Boltzmanngasse 5, 1090 Vienna, Austria}

\author{Lixiang Chen}
\email{chenlx@xmu.edu.cn}
\affiliation{Department of Physics and Collaborative Innovation Center for Optoelectronic Semiconductors and Efficient Devices, Xiamen University, Xiamen 361005, China}

\author{Fabian Steinlechner}
\email{Fabian.Steinlechner@iof.fraunhofer.de}
\affiliation{Fraunhofer Institute for Applied Optics and Precision Engineering IOF, Albert-Einstein-Strasse 7, 07745 Jena, Germany.}
\affiliation{Friedrich Schiller University Jena, Abbe Center of Photonics, Albert-Einstein-Str. 6, 07745 Jena, Germany.}

\author{Marcus Huber}
\affiliation{
Institute for Quantum Optics and Quantum Information - Vienna (IQOQI), Austrian Academy of Sciences, Boltzmanngasse 3, 1090 Vienna, Austria.}

\author{Rupert Ursin}
\email{Rupert.Ursin@oeaw.ac.at}
\affiliation{
Institute for Quantum Optics and Quantum Information - Vienna (IQOQI), Austrian Academy of Sciences, Boltzmanngasse 3, 1090 Vienna, Austria.}
\affiliation{Vienna Center for Quantum Science \& Technology (VCQ), Faculty of Physics, University of Vienna, Boltzmanngasse 5, 1090 Vienna, Austria}


\begin{abstract}
High-dimensional quantum entanglement is currently one of the most prolific fields in quantum information processing due to its high information capacity and error resilience. A versatile method for harnessing high-dimensional entanglement has long been hailed as an absolute necessity in the exploration of quantum science and technologies. Here we exploit Hong-Ou-Mandel interference to manipulate discrete frequency entanglement in arbitrary-dimensional Hilbert space. The generation and characterization of two-, four- and six-dimensional frequency entangled qudits are theoretically and experimentally investigated, allowing for the estimation of entanglement dimensionality in the whole state space. Additionally, our strategy can be generalized to engineer higher-dimensional entanglement in other photonic degrees of freedom. Our results may provide a more comprehensive understanding of frequency shaping and interference phenomena, and pave the way to more complex high-dimensional quantum information processing protocols.
\end{abstract}

\pacs{}

\maketitle

\indent \emph{Introduction.}\rule[2pt]{8pt}{1pt}Harnessing entanglement in high-dimensional systems may well play a central role in elevating the performance of advanced quantum information protocols towards practical applicability. In particular in the context of quantum communication, photon pairs entangled in high dimensions can carry more quantum information, making them compelling for enhancing quantum channel capacities, improving noise resilience and even speeding up certain tasks in photonic quantum computation \cite{PhysRevX.9.041042,dada2011experimental,mirhosseini2015high,raussendorf2007fault}. Several physical properties of photons can be used to directly encode high-dimensional entanglement, including orbital angular momentum \cite{molina2007twisted}, time-energy \cite{zhong2015photon} and path \cite{fickler2014interface}. The drawback of employing high-dimensional encoding in the spatial domain is the stringent requirement on the quality of optical wave-fronts and shaping for generation and measurement, hence confining its applications in optical fiber based quantum communication systems \cite{kovlakov2017spatial}. Conversely, entanglement in the energy-time domain is intrinsically suitable for long-distance transmission in fiber and free space \cite{Steinlechner2017}. Among the many manifestations of energy-time entanglement, photons carrying frequency entanglement have attracted great interest in recent years \cite{kuzucu2005two,roslund2014wavelength,yan2011generation,lingaraju2019quantum,venkata2019interfernce}. However, the post-selection-free creation of arbitrary-dimensional frequency entanglement still remains relatively unexplored. In addition, the characterization and verification of high-dimensional frequency entanglement poses an ongoing challenge, owing to both the difficulty of performing required superposition measurements in the frequency domain, as well as the general challenges associated with performing full quantum state tomography in a large state space.

The objective of this work is twofold: Firstly, we use spatial beating of Hong-Ou-Mandel (HOM) interference with polarization-frequency hyperentangled photons on a beam-splitter to discretize continuous and broadband spectra into a series of frequency bins. Then, we exploit polarization anti-correlations of hyperentangled state to deterministically eliminate noise contributions that arise from unwanted photon bunching in the preparation stage, thus achieving high-fidelity frequency-bin entanglement without the usual requirement of detection post-selection. Secondly, we show how HOM interference can be used to characterize high-dimensional frequency entanglement; measurements of the fringe spacing of the observed interference pattern allow us to extract individual parameters even for high-dimensional entanglement. These results demonstrate that the modulation of temporal distinguishability in HOM interference provides a powerful tool for implementing high-dimensional quantum information processing.


\indent \emph{Discretization of continuous frequency entanglement.}\rule[2pt]{8pt}{1pt}
\begin{figure*}[!t]
\centering
\subfigure{
\includegraphics[width=0.96\linewidth]{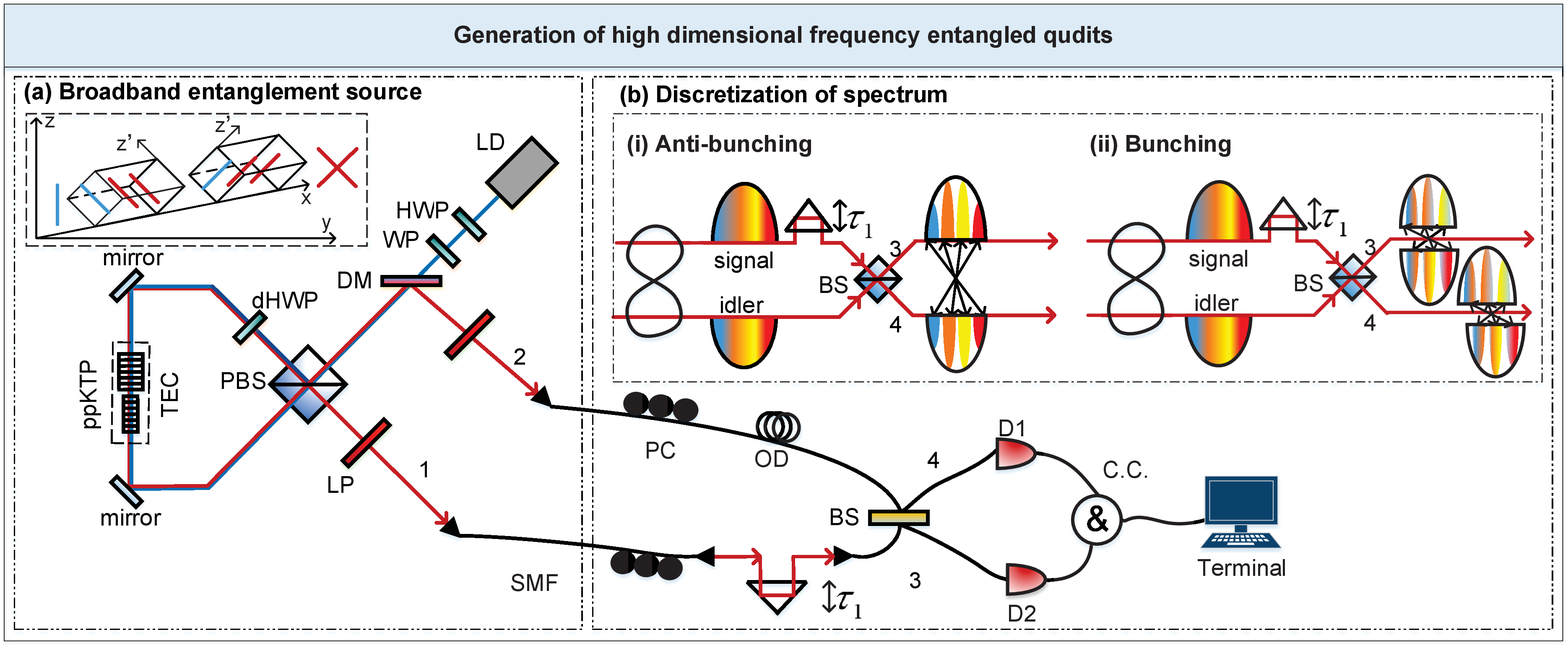}}
\subfigure{
\includegraphics[width=0.48\linewidth]{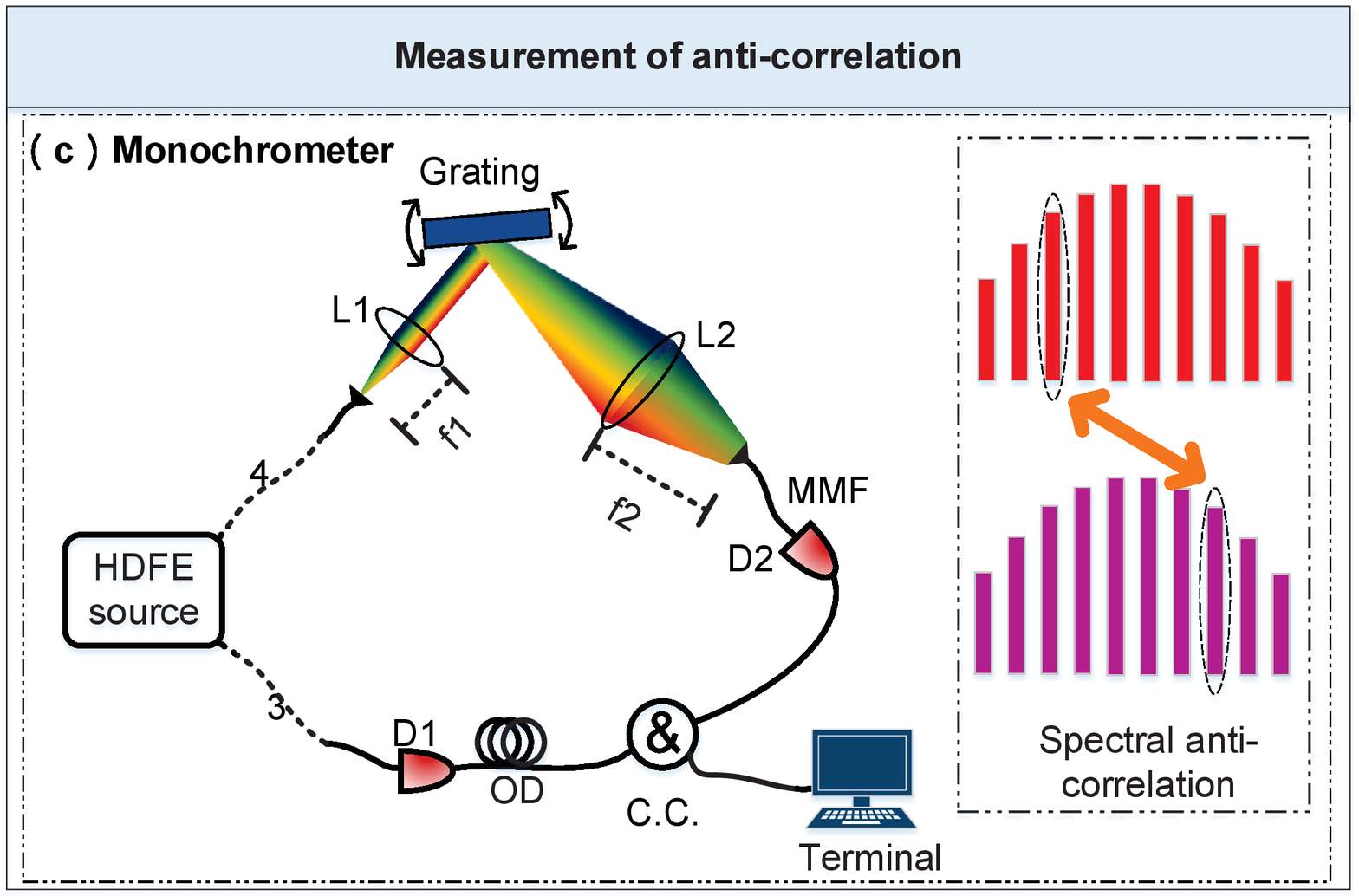}}
\subfigure{
\includegraphics[width=0.48\linewidth]{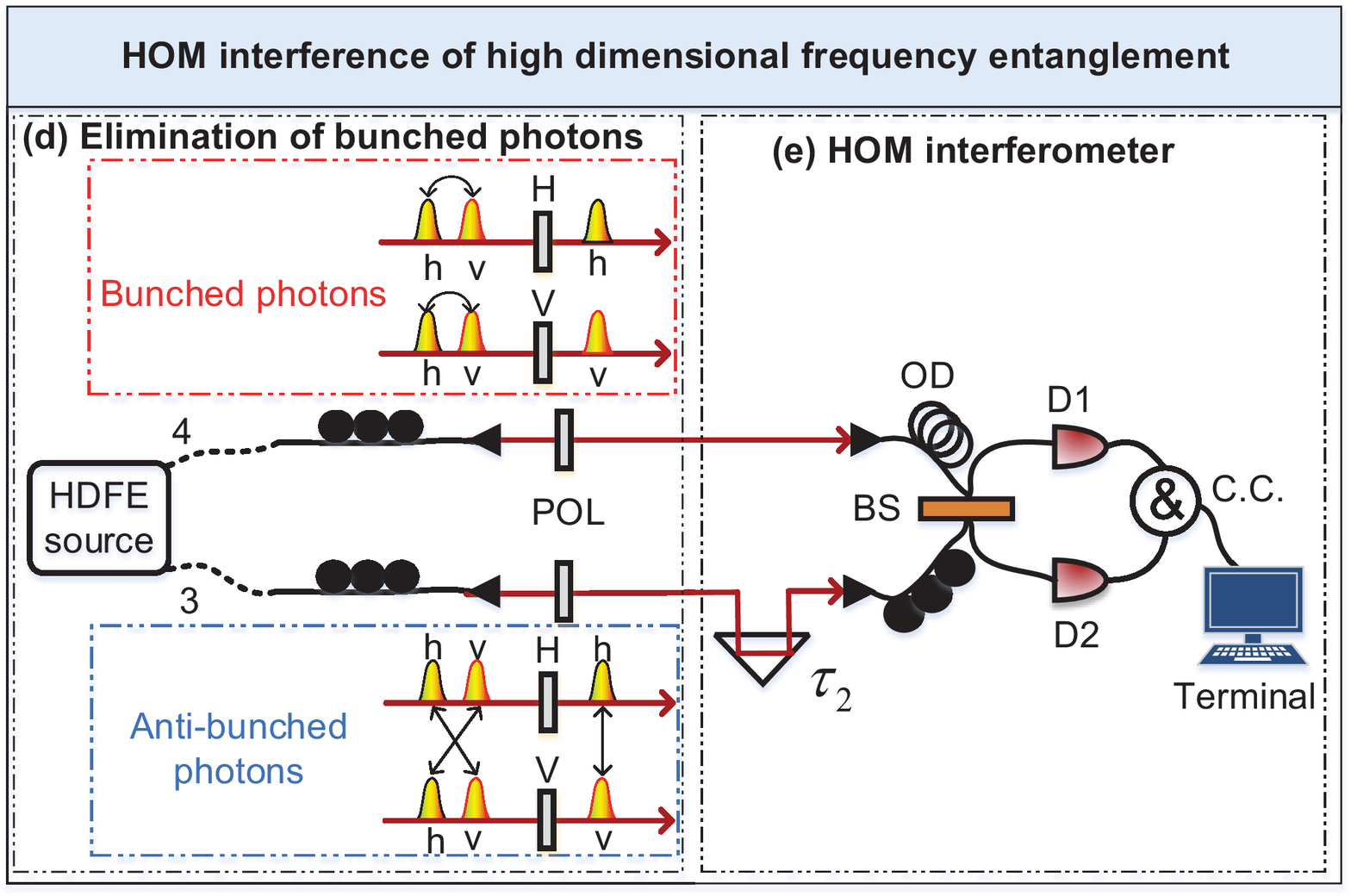}}
\caption{(a) Generation of high-dimensional frequency-entangled qudits through (b) Hong-Ou-Mandel interference. (c) Observation of frequency anti-correlations with a monochromator. Assisted by the elimination of spatially bunched photons (d), HOM interference of high dimensional entanglement is observed (e). LD: laser diode; PBS: polarizing beam splitter; HWP: half wave plate; DM: dichroic mirror; TEC: temperature controller; LP: long pass filter; SMF: single-mode fiber; BS: beam splitter; MMF: multi-mode fiber; POL: polarizer.
}
\label{figure_1}
\end{figure*}
A major limiting factor for the number of entangled frequency modes, quantified by the dimensionality of entanglement, is the spectral bandwidth of photons. Type-0 phase-matching yields the broadest spectral bandwidths among spontaneous parametric down-conversion (SPDC) processes \cite{soren2018entanglement,fabian2012high}. By combining the benefits of Sagnac-type sources \cite{fedrizzi2007wavelength} and crossed-crystal sources \cite{kwiat1999ultrabright}, wavelength-degenerate polarization-entangled photons are deterministically routed into two distinct spatial modes (see Fig. \ref{figure_1}\textcolor{blue}{(a)}) as a direct consequence of time-reversed HOM interference \cite{chen2018polarization}.  The resulting  polarization-entangled state can be written in the form of
$\ket{\psi}=\frac{1}{\sqrt{2}}(\ket{H_1V_2}+\ket{V_1H_2})\otimes\ket{\omega_s\omega_i}$, where H (V) denotes horizontal (vertical) polarization, and $\omega_s$ and $\omega_i$ denote the continuous spectrum of signal and idler photons. Since the difference frequency $\Delta\omega=|\omega_s-\omega_i|$ of a color-entangled state typically exceeds that of the pump laser, continuous entanglement arises quite naturally as a consequence of energy conservation.
\begin{figure}[!t]
\centering
\subfigure[]{
\label{Fig2.sub.1}
\includegraphics[width=0.48\linewidth]{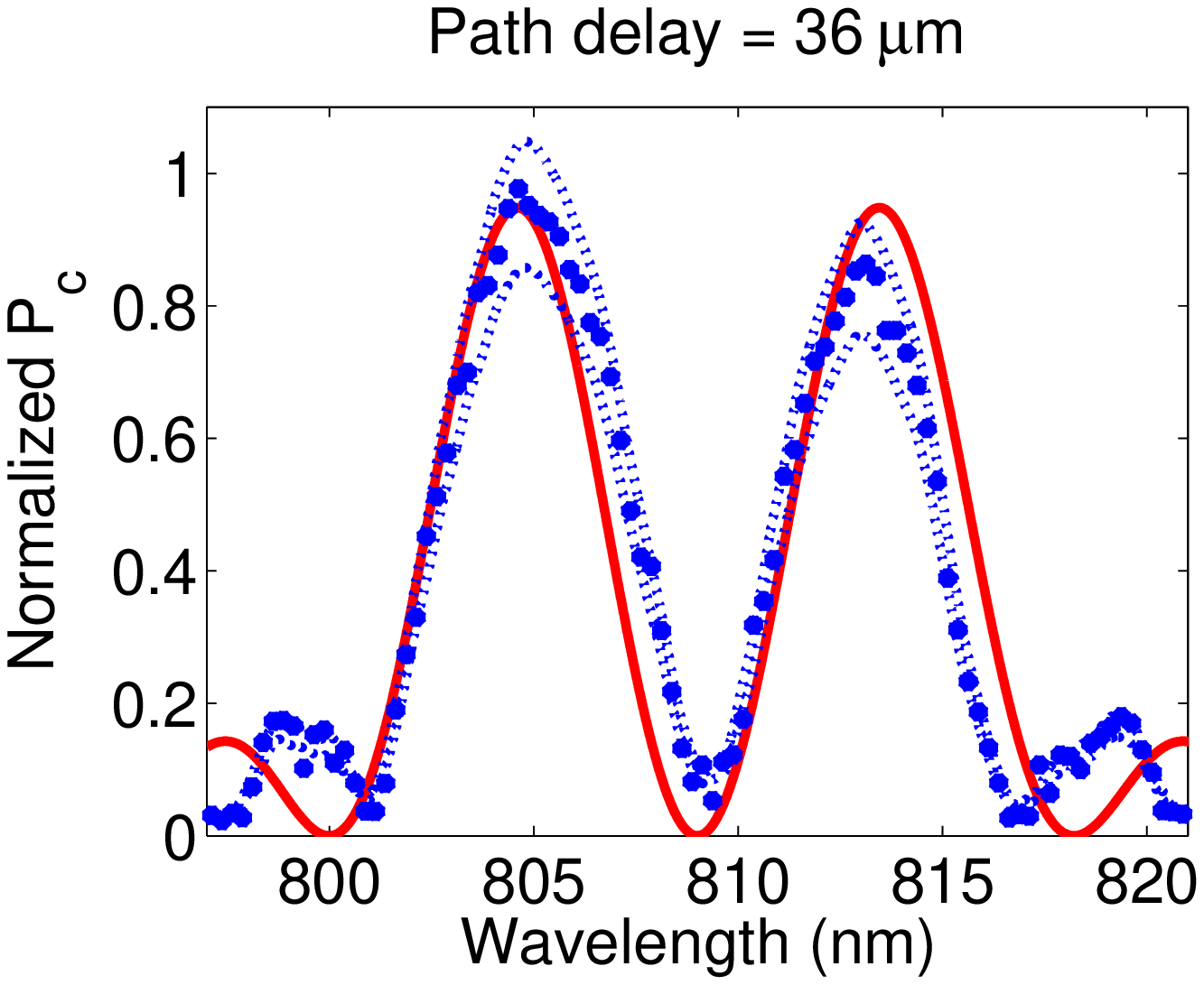}}
\subfigure[]{
\label{Fig2.sub.2}
\includegraphics[width=0.48\linewidth]{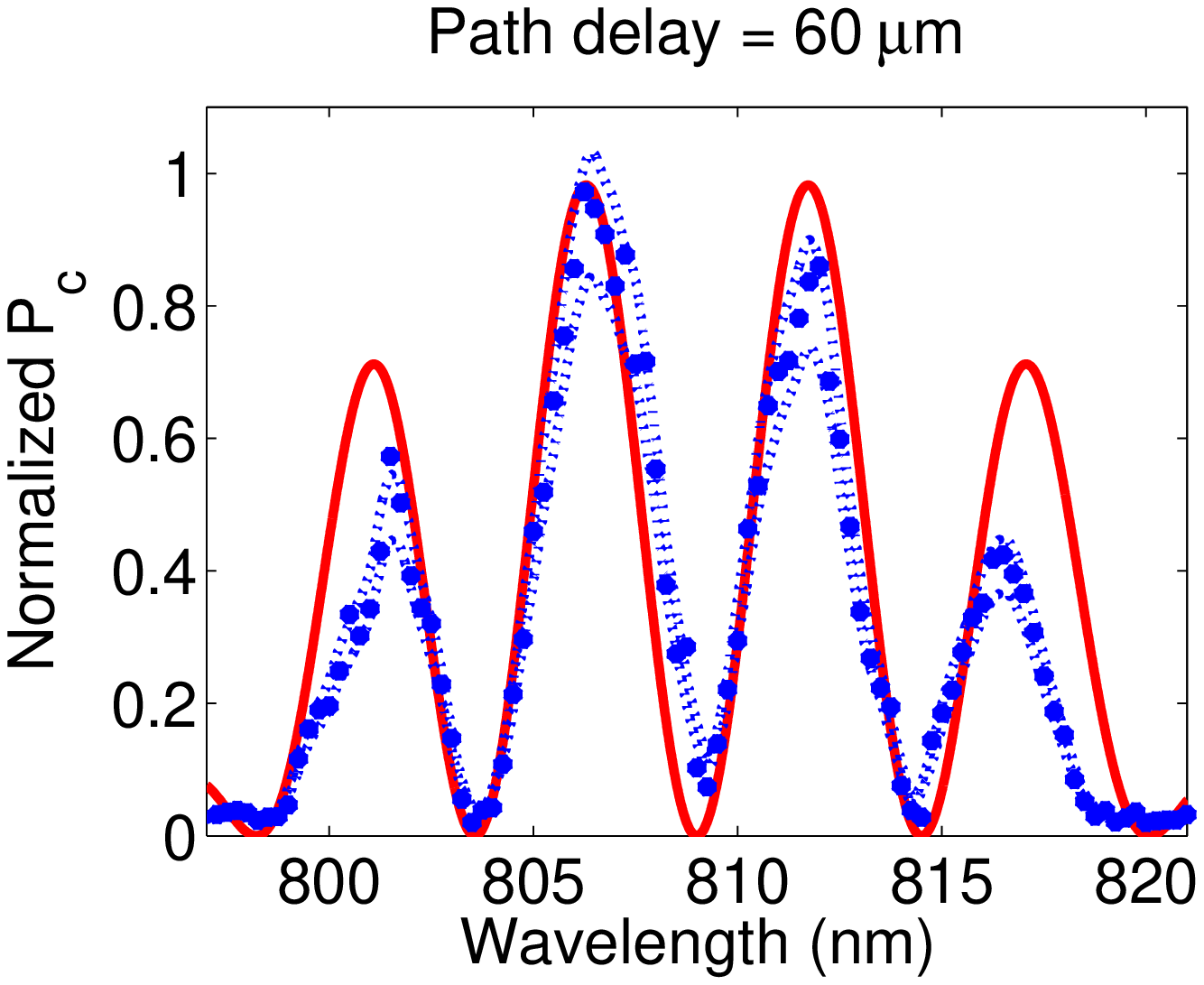}}
\subfigure[]{
\label{Fig2.sub.3}
\includegraphics[width=0.48\linewidth]{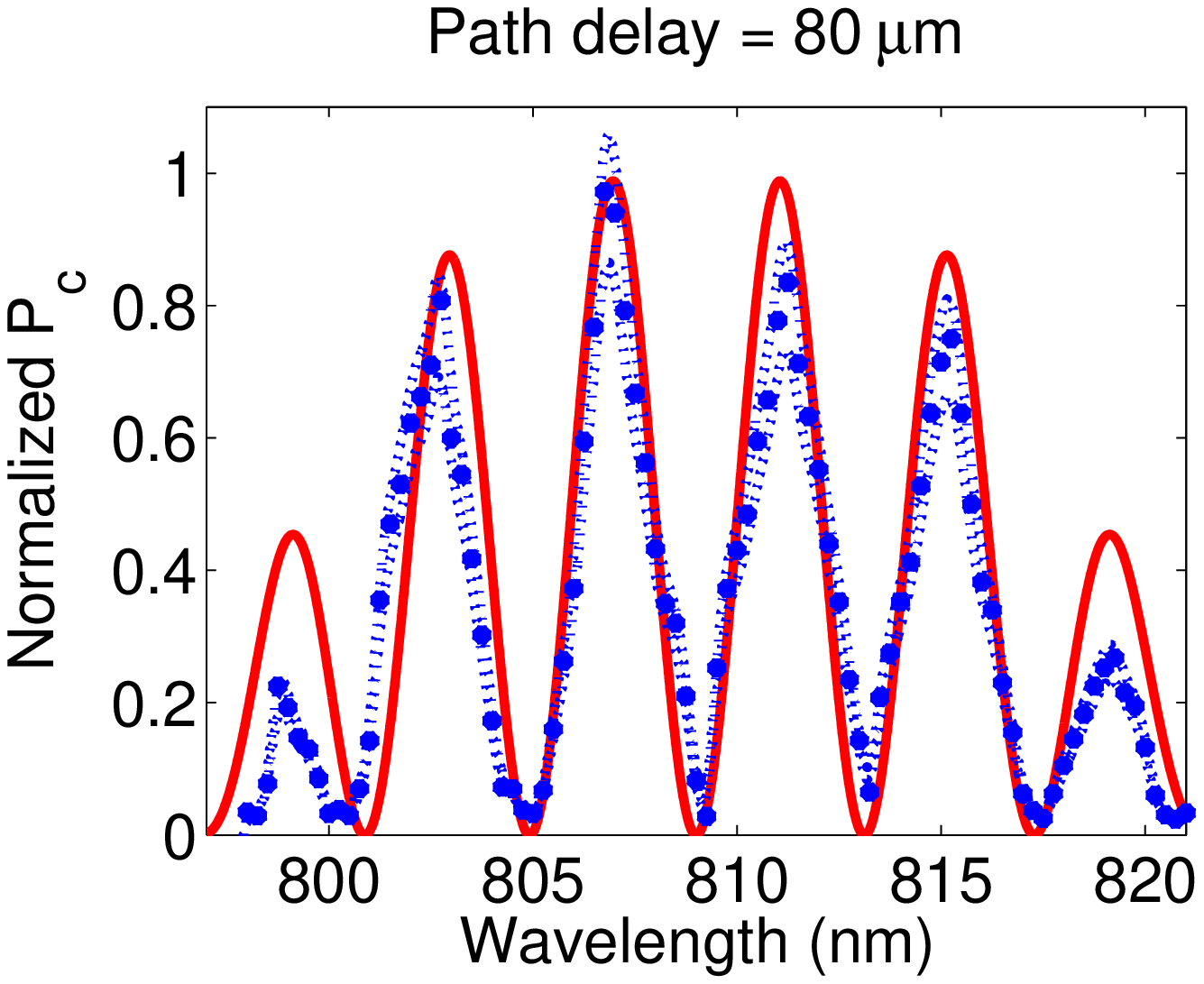}}
\subfigure[]{
\label{Fig2.sub.4}
\includegraphics[width=0.48\linewidth]{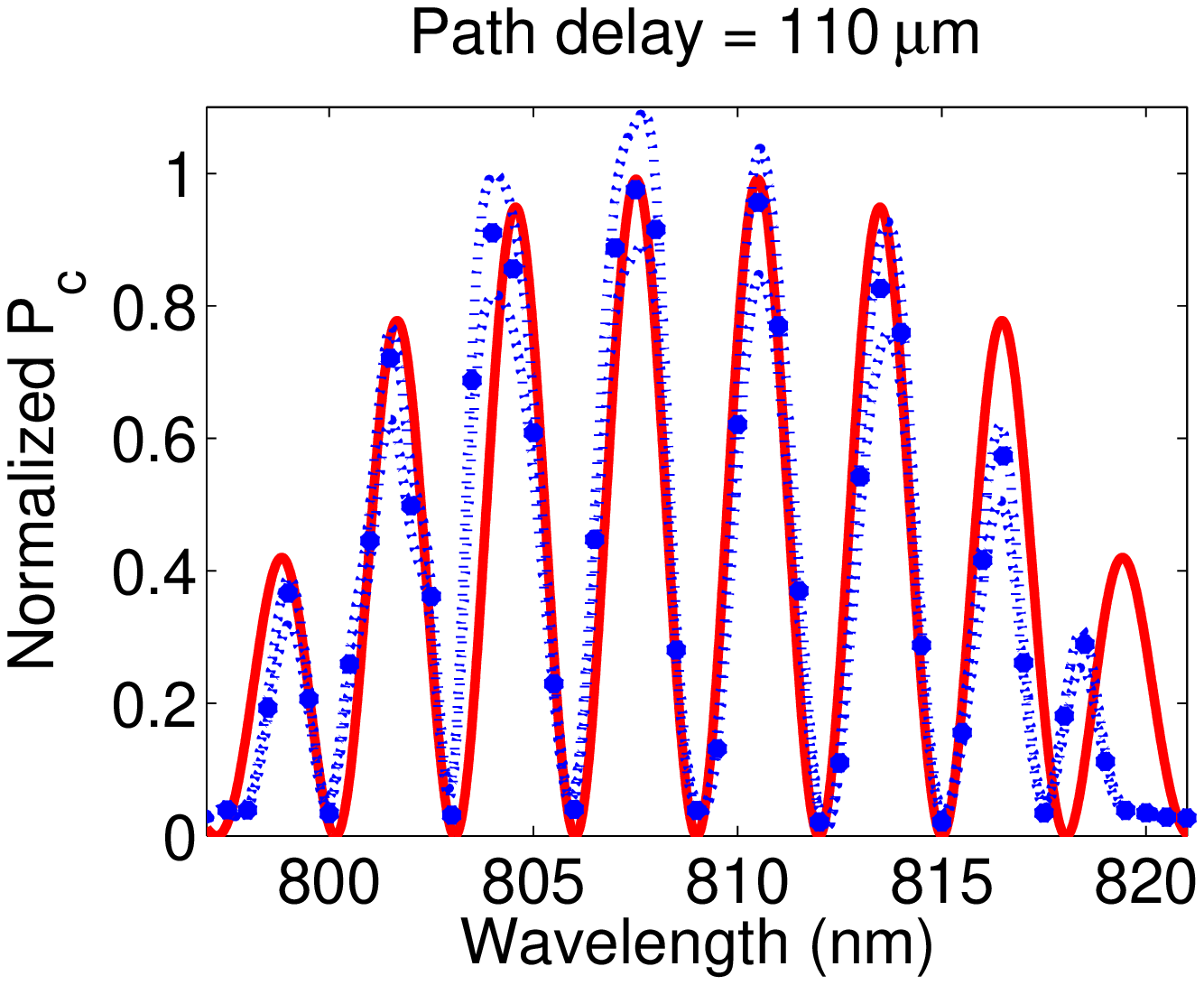}}
\caption{Observation of frequency anti-correlation by setting the relative path (time) delay at (a) \unit[36]{$\mu$m} ($\tau_1 = \unit[0.12]{ps}$), (b) \unit[60]{$\mu$m} ($\tau_1 = \unit[0.2]{ps}$) (c) \unit[80]{$\mu$m} ($\tau_1 = \unit[0.27]{ps}$) and (d) \unit[110]{$\mu$m} ($\tau_1 = \unit[0.37]{ps}$). The red lines represent the theoretical predictions, while the blue points represent the experimental results that are bounded by the standard deviation estimated by statistical methods assuming a Poisson distribution.}
\label{figure_2}
\end{figure}
Then, a HOM interferometer is used to project the continuous frequency spectrum into well-separated frequency bins without any requirement for spectrally selective filtering. As depicted in Fig. \ref{figure_1}\textcolor{blue}{(b)}, the photon in path 1 is then delayed relative to its partner photon in path 2, introducing a relative time delay $\tau_1$ between the photons' time-of-arrival on a balanced beam splitter. The bi-photon state after this beam splitter is transformed to
\begin{equation}
\begin{split}
\ket{\psi}_{\text{hyper}}= &\frac{1}{2}[(\ket{H_3V_4}+\ket{V_3H_4})\otimes\ket{\psi}_\omega^-\\
                   &+(\ket{H_3V_3}+\ket{H_4V_4})\otimes\ket{\psi}_\omega^+],
\end{split}
\label{eq:hyper}
\end{equation}
where $\ket{\psi}_\omega^-$ ($\ket{\psi}_\omega^+$) denotes the frequency-entangled state produced in opposite (identical) spatial modes \cite{jin2016simple}. The drawback of this transformation is that the bi-photon component $\ket{\psi}_\omega^+$ diminishes the interference visibility in the characterization process. To tackle this issue, we eliminate the detrimental bunched photon events via harnessing the anti-correlations of polarization-entangled state (see Fig.\ \ref{figure_1}\textcolor{blue}{(d)}). This can be considered a state purification in the frequency domain and leads to a single non-vanishing term $\ket{V_3H_4}\otimes\ket{\psi}_\omega^-$. By identifying photons in opposite spatial modes, the normalized coincidence probability can be
\begin{equation}\label{eq: pt1}
\begin{split}
P(\tau_1)=&\frac{1}{4}\int\int d\omega_1d\omega_2f(\omega_1,\omega_2)|1-e^{i(\omega_2-\omega_1)\tau_1}|^2.
\end{split}
\end{equation}
This indicates that photon anti-bunching, i.e., coincidence events, only occur when bi-photons' frequency detuning satisfies $(\omega_1-\omega_2)\tau_1=(2n+1)\pi$ ($n$ is an integer). The intensity of bi-photon component in opposite spatial modes manifests itself as a sinusoidal oscillation within a Gaussian envelope, with an oscillation period related to $\tau_1$. As a consequence, arbitrary-dimensional discrete frequency entanglement can be prepared by tuning the temporal distinguishability of frequency-entangled photons in HOM interference.

The joint spectral amplitude for different $\tau_1$ are plotted in Fig. \ref{figure_2} as a function of signal and idler wavelengths. The created discrete frequency-entangled state can be
\begin{equation}
\begin{split}
\ket{\psi}=\sum_{j=1}^{m/2}A_j(\alpha_j\ket{\omega_j\omega_{m-j}}-e^{i\varphi_j}\alpha_{m-j}\ket{\omega_{m-j}\omega_j}),
\end{split}
\label{eq:4}
\end{equation}
where $m$ denotes the number of dimensions, $A_j$ is a probability amplitude, $\varphi_j$ is a phase-offset, $\alpha_j^2=p_j$, $\alpha_{j}^2+\alpha_{m-j}^2=1$, $p_j$ is a balance parameter and $\omega_j+\omega_{m-j}=\omega_p$. We note that the frequency bins are completely symmetric with respect to the central frequency. This agrees well with the fact that when wavelength-degenerate photons impinge on a beam splitter simultaneously, the coincidence probability falls to zero (so-called ``HOM dip'') \cite{hong1987measurement}, since all distinguishing information of the photons is erased.

\indent \emph{Observation of frequency anti-correlation.}\rule[2pt]{8pt}{1pt}
Polarization-entangled photon pairs are created in a Sagnac-type source (see Fig. \ref{figure_1}\textcolor{blue}{(a)}) at a degenerate wavelength of $\unit[810]{nm}$ with a spectral bandwidth of $\sim\unit[20]{nm}$. They are produced at a pair rate of 160 kcps per mW of pump power with a fidelity to the Polarization Bell state of 99.2\% \cite{chen2018polarization}. The existence of multiple well-separated frequency bins is experimentally verified using a home-made single-photon monochromator (see Fig.\ \ref{figure_1}\textcolor{blue}{(c)}). By identifying two-fold coincidences in opposite spatial modes 3 and 4, we experimentally investigate anti-correlations of frequency-entangled qudits by setting the relative time delays at $\unit[0.12]{ps}$, $\unit[0.2]{ps}$, $\unit[0.27]{ps}$ and $\unit[0.37]{ps}$. The experimental results are depicted in Fig.\ \ref{figure_2}, which indicates a good separation of multiple frequency modes, and allows us to quantify the corresponding parameters as shown in Table \ref{tab1}.
\begin{table}[!t]
\begin{tabular}{p{0.8cm}<{\centering}|p{1.7cm}<{\centering}|p{1.8cm}<{\centering}|p{1.8cm}<{\centering}|p{1.8cm}<{\centering}}
\hline
\hline
$\tau_1$(ps) & $\Delta\lambda_{\text{FWHM}}$(nm) & $\mu_j$ (THz) & $p_j$ & $A_j$\\
\hline
0.12 & 4.33 & 3.93 & 0.54 & -\\
\hline
0.20 & 2.83 & 6.94/2.67 & 0.57/0.53 & 0.36/0.65\\
\hline
0.27 & 2.11 & 5.71/1.94 & 0.52/0.54 & 0.46/0.54\\
\hline
0.37 & 1.52 & 6.91/4.09/1.42 & 0.56/0.52/0.50 & 0.24/0.35/0.41\\
\hline
\end{tabular}
\caption{Parameters extracted from Fig. \ref{figure_2}. $\Delta\lambda_{\text{FWHM}}$ is the single-photon frequency bandwidth, $\mu_j=|\omega_j-\omega_{m-j}|$ is the frequency detuning.}
\label{tab1}
\end{table}

\begin{figure}[!t]
\centering
\subfigure[]{
\label{Fig3.sub.1}
\includegraphics[width=0.48\linewidth]{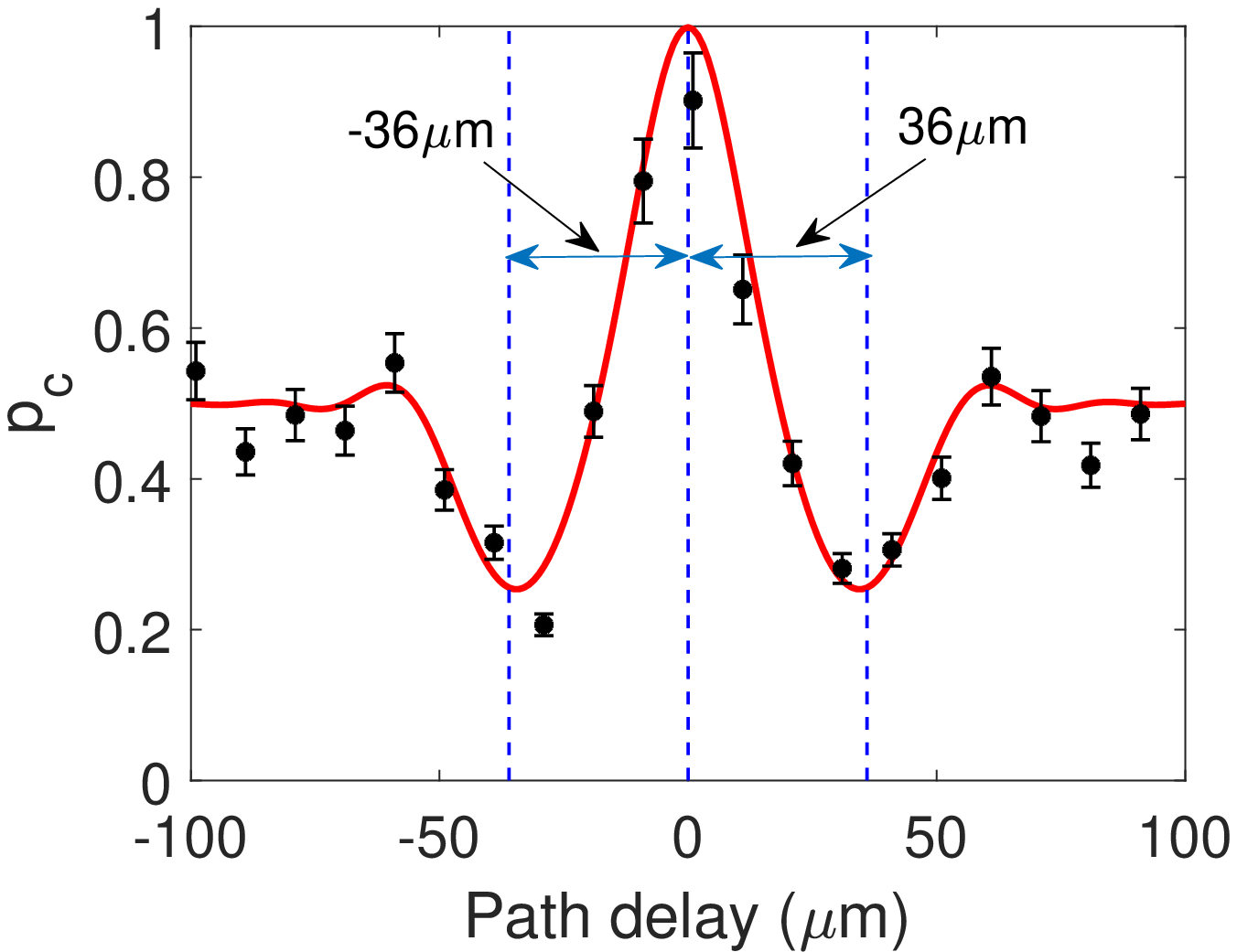}}
\subfigure[]{
\label{Fig3.sub.2}
\includegraphics[width=0.48\linewidth]{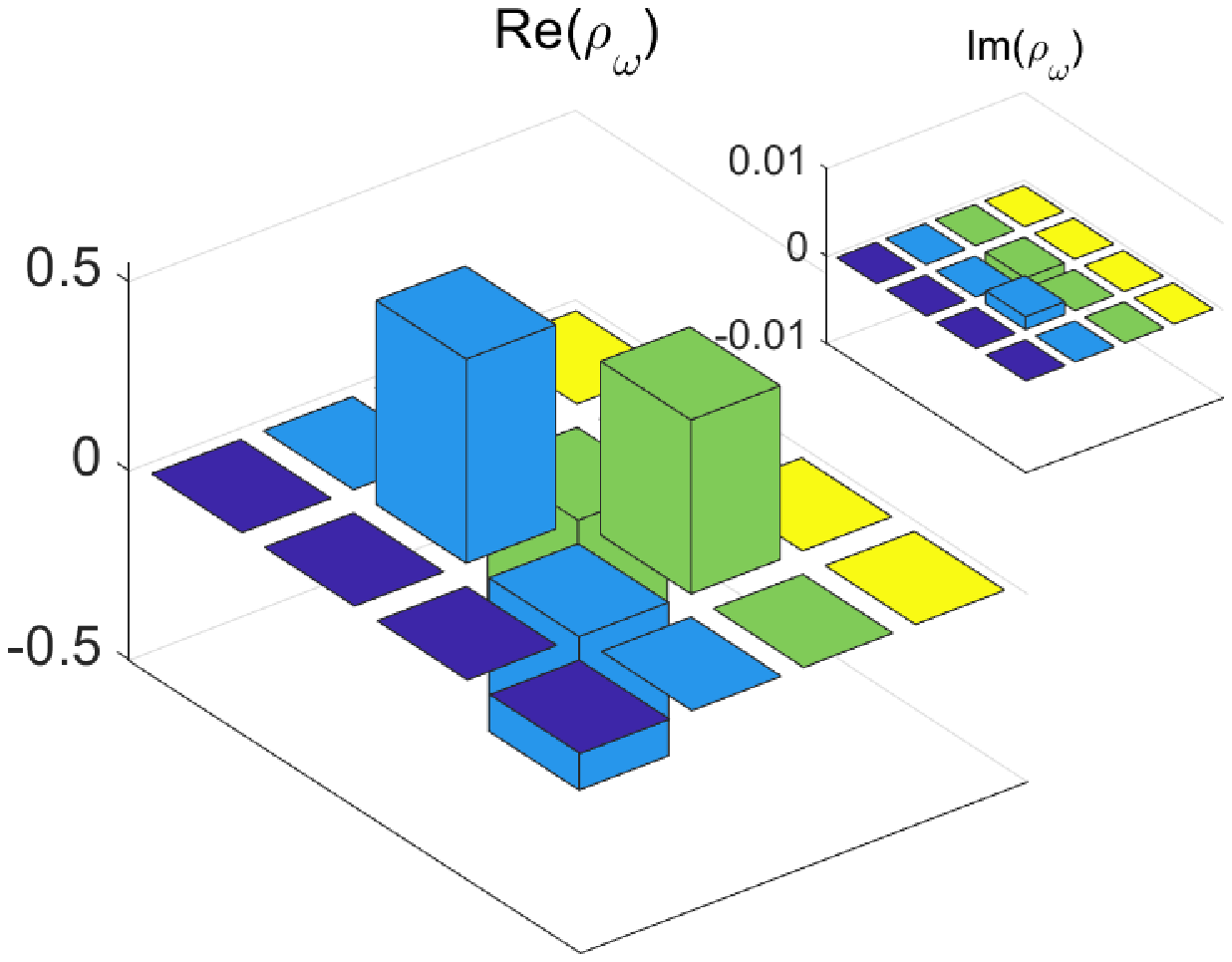}}
\subfigure[]{
\label{Fig3.sub.3}
\includegraphics[width=0.48\linewidth]{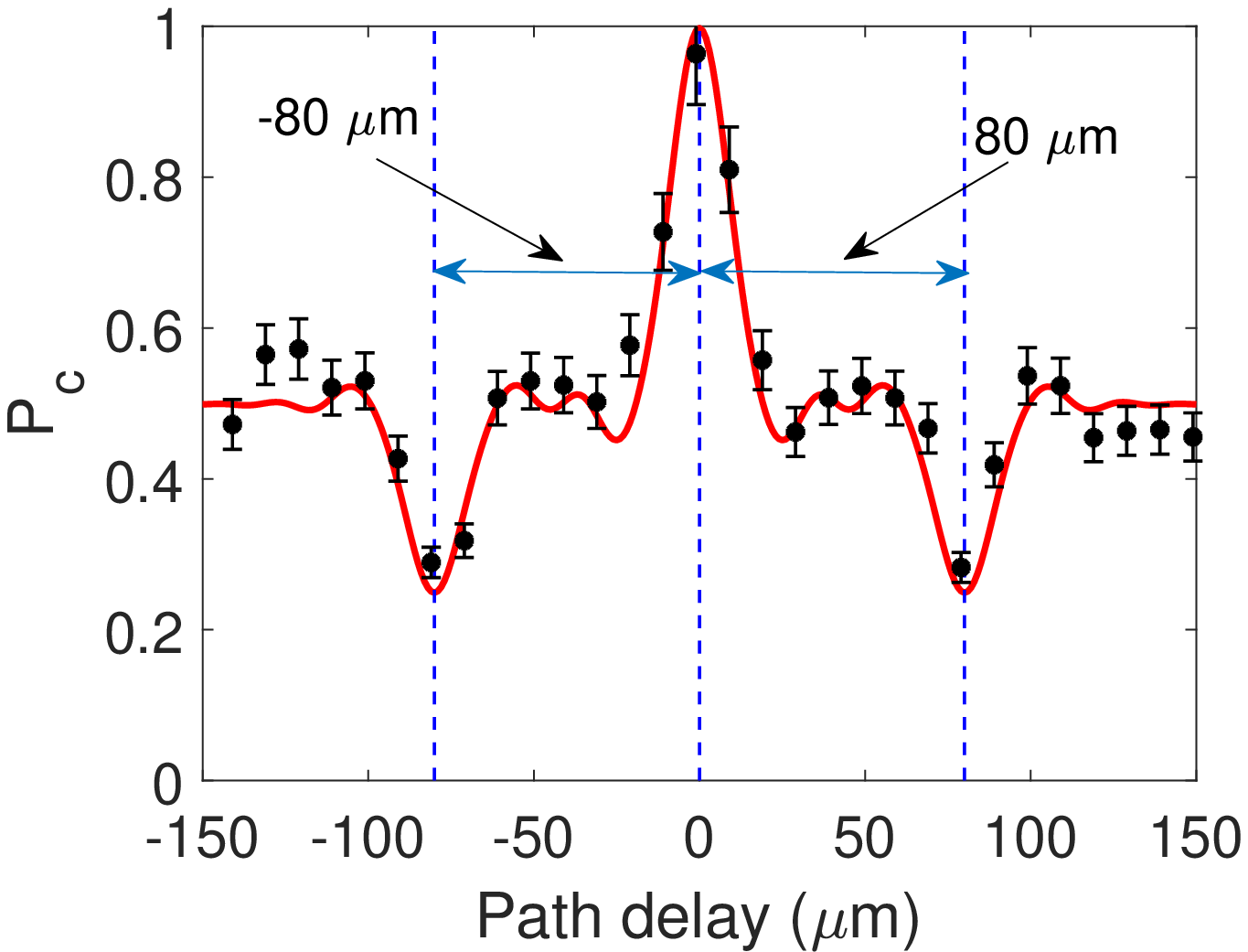}}
\subfigure[]{
\label{Fig3.sub.4}
\includegraphics[width=0.48\linewidth]{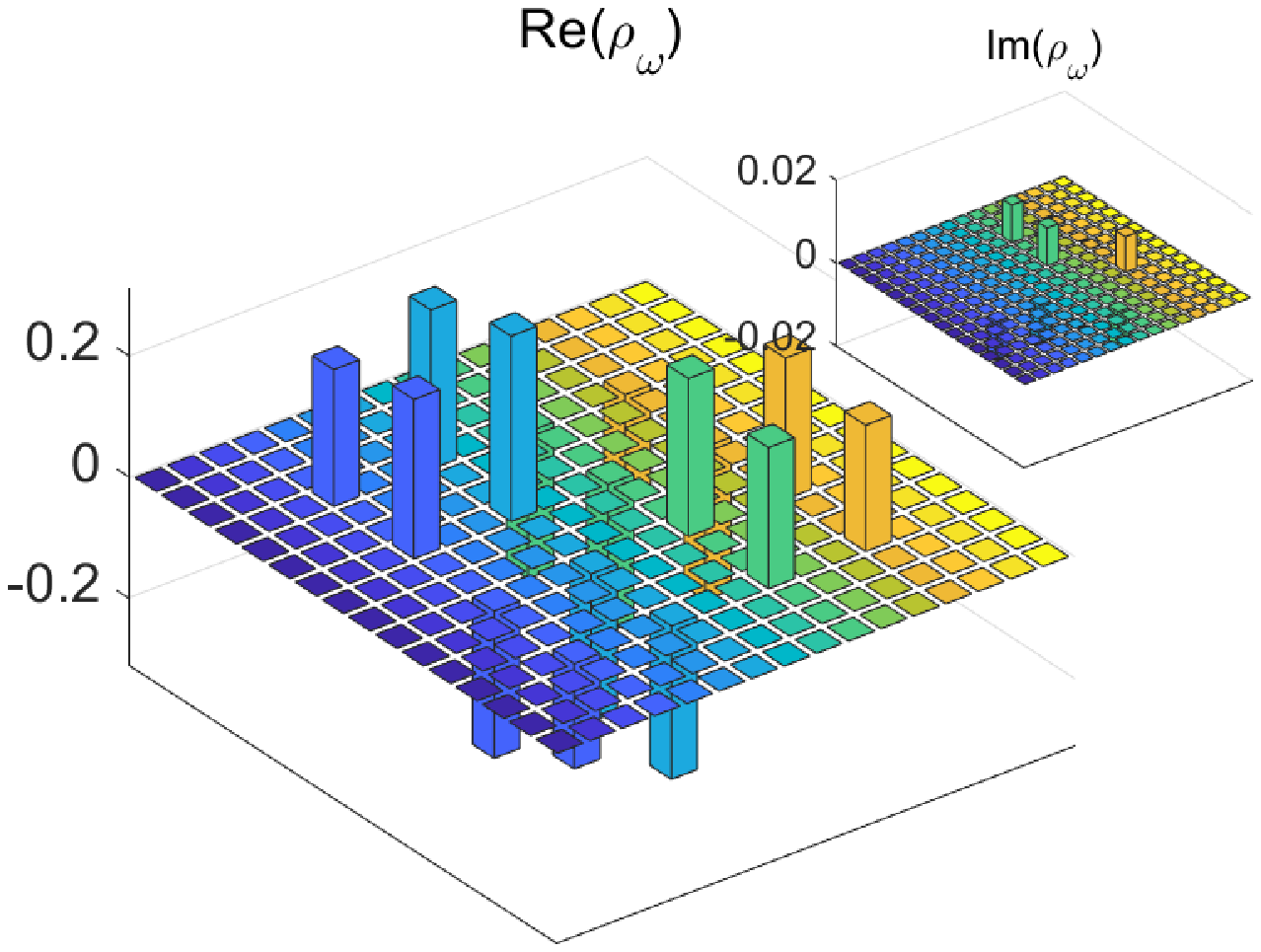}}
\subfigure[]{
\label{Fig3.sub.5}
\includegraphics[width=0.48\linewidth]{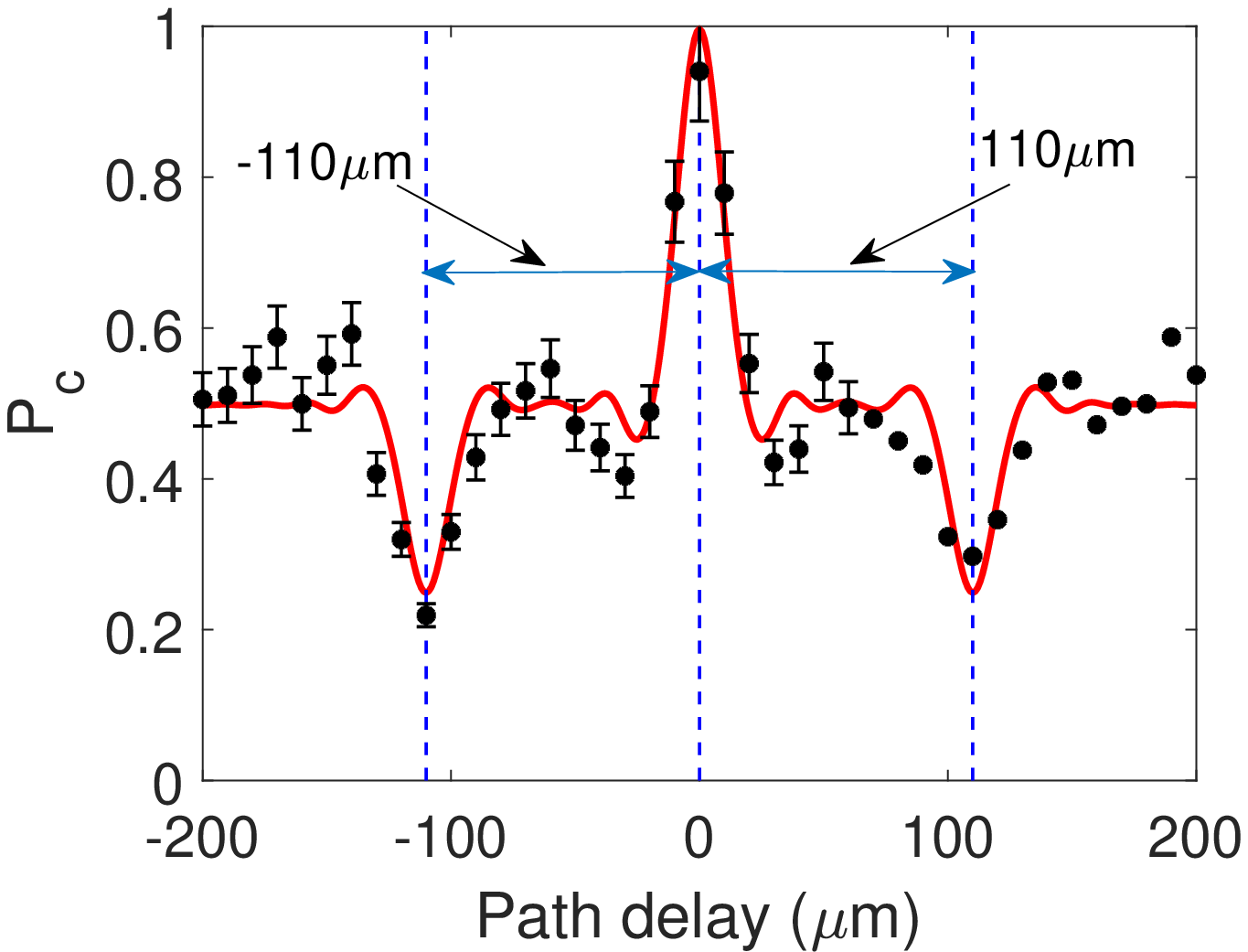}}
\subfigure[]{
\label{Fig3.sub.6}
\includegraphics[width=0.48\linewidth]{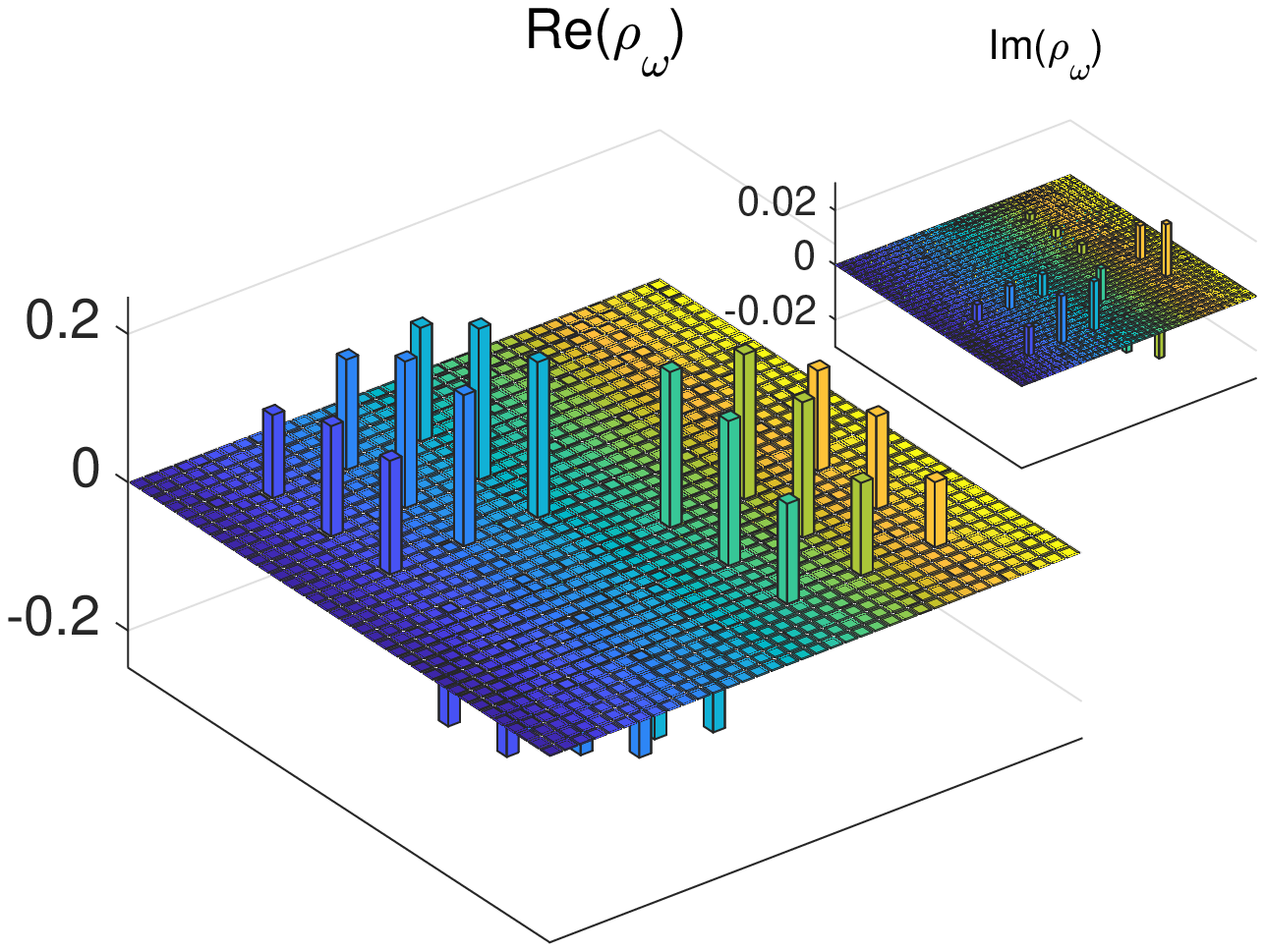}}
\caption{Spatial beating of discrete frequency-entangled photon pairs of (a-b) two-dimensional frequency entanglement when setting $\tau_1 = \unit[0.12]{ps}$ (\unit[36]{$\mu$m}), (c-d) four-dimensional frequency entanglement when setting $\tau_1 = \unit[0.27]{ps}$ (\unit[80]{$\mu$m}) and (e-f) six-dimensional frequency entanglement when setting $\tau_1 = \unit[0.37]{ps}$ (\unit[110]{$\mu$m}). (a,c,e) Theoretical prediction and experimental results of the normalized coincidence rate, (b,d,f) real and imaginary parts of the restricted density matrix.}
\label{figure_3}
\end{figure}

\indent \emph{HOM interference of high-dimensional frequency entanglement.}\rule[2pt]{8pt}{1pt}
The characterization of frequency entanglement is challenging since the direct observation of interference visibilities in superposition bases of discrete frequency bins is a non-trivial task without nonlinear optics or time-resolved measurements \cite{kues2017chip, Lu2018,Maclean2018,Brecht2015}. Here, we implement a non-local measurement of the coherences of a frequency-entangled state by spatial beating in HOM interference \cite{ou1988observation} (see Fig.\ \ref{figure_1}\textcolor{blue}{(d)}). The imbalance between two arms of a HOM interferometer introduces a relative time delay $\tau_2$, thus the coincidence probability $P(\tau_1,\tau_2)$ is obtained as
\begin{equation}\label{eq:second HOM}
\begin{split}
P(\tau_1,\tau_2)=&\frac{1}{16}\int\int d\omega_1d\omega_2f(\omega_1,\omega_2)\\
&|e^{-i\omega_2(\tau_1+\tau_2)}+e^{-i(\omega_1\tau_1+\omega_2\tau_2)}\\
&+e^{-i\omega_1(\tau_1+\tau_2)}e^{-i(\omega_2\tau_1+\omega_1\tau_2)}|^2.
\end{split}
\end{equation}
The parameter $\tau_1$ is set to be a constant that effectively determines the entanglement dimensionality, thus $P(\tau_1,\tau_2)$ is simplified to the interference probability over the spatial beating delay $\tau_2$. Figures \ref{figure_3}\textcolor{blue}{(a,c,e)} are the theoretical results of \eqref{eq:second HOM} by setting $\tau_1$ to be $\unit[0.12]{ps}$, $\unit[0.27]{ps}$ and $\unit[0.37]{ps}$, where two-, four- and six-dimensional frequency entanglement are prepared, respectively. The manifestation of interference fringes can be approximated by the sum of coincidence probabilities with different frequency detunings as \cite{fedrizzi2009anti}
\begin{equation}
\label{eq:P2}
\begin{split}
P(\tau_2)=\frac{1}{2}-\sum_{j=1}^{m/2}\frac{V_j}{2}A_jcos(\mu_j\tau_2+\varphi_j)(1-|\frac{2\tau_2}{\tau_c}|),
\end{split}
\end{equation}
for $|\tau_2|\leq \tau_c/2$, where $V_j$ is the interference visibility quantifying the magnitude of oscillation, $\mu_j=|\omega_j-\omega_{m-j}|$ is frequency detuning of wellseparated frequency-entangled bins, and $\tau_c$ denotes the single-photon coherence time that equals to the base-to-base envelope width. Curve fitting of experimental results to \eqref{eq:P2} reveal the corresponding parameters of a restricted density matrix in the high-dimensional state space.

\indent \emph{Experimental characterization of high-dimensional frequency entanglement.}\rule[2pt]{8pt}{1pt}
We utilize two polarizers oriented at mutually orthogonal settings to eliminate the photon bunching components in \eqref{eq:hyper}, as shown in the insets of Fig.\ \ref{figure_1}\textcolor{blue}{(d)}. In addition, polarization controllers are used to erase any polarization distinguishability. Then, the photons from each pair impinge on a balanced beam splitter from distinct input ports (see Fig.\ \ref{figure_1}\textcolor{blue}{(e)}). The non-classical spatial beating is revealed by scanning the arrival time of one of the photons incident on the beam splitter by $\tau_2$. The resulting interference fringes can be observed by identifying two-fold coincidence events between opposite output ports of the beam splitter.

The experimental interference fringes for two-, four- and six-dimensional frequency entanglement created with delays $\tau_1 = \unit[0.12]{ps}$, $\unit[0.27]{ps}$ and $\unit[0.37]{ps}$ are depicted in Fig. \ref{figure_3} \textcolor{blue}{(a,c,e)}. Table \ref{tab2} shows the corresponding parameters extracted from experimental measurement by using nonlinear curve-fitting. The parameters enable us to reconstruct the restricted density matrices as shown in Fig. \ref{figure_3} \textcolor{blue}{(b,d,f)}. We note that our experimental results agree well with theoretical prediction, while slight deviations can be attributed to misalignment and imperfect optical components.
\begin{table}[!t]%
\begin{tabular}{p{0.5cm}<{\centering}|p{0.5cm}<{\centering}|p{1.1cm}<{\centering}|p{1.15cm}<{\centering}|p{1.15cm}<{\centering}|p{1.15cm}<{\centering}|p{1.7cm}<{\centering}}
\hline
\hline
$\tau_1$ (ps) &$\tau_c$ (ps) &$\Delta\lambda_{\text{FWHM}}$ (nm) & $\mu_j$ (THz) & $A_j$ & $V_j$ & $\varphi_j$($^\circ$)\\
\hline
0.12            &0.47      & 4.24 & 4.01        & -              & 0.81           & 179.83\\
\hline
0.27          & 0.94        & 2.13 & 5.71/1.94   & 0.44/0.56       &   0.80/0.86    & 180.03/182.14\\
\hline
0.37          & 1.43        & 1.42 & 6.54/4.07 & 0.20/0.38  & 0.80/0.94  & 172.48/181.23\\
~             & ~        & ~    & /1.48     & /0.42      & /0.93      &  /177.67 \\
\hline
\end{tabular}
\caption{Parameters extracted from Fig \ref{figure_3}.}
\label{tab2}
\end{table}

On the basis of previous discussion, the HOM interference dip appears when $\tau_2$ is odd times of $\tau_1$, while peak appears when $\tau_2$ is even times of $\tau_1$. Additionally, since the spectrum manifests itself as a characteristic sinusoidal function, the single photon frequency bandwidth can be approximated as $\Delta f_{\text{FWHM}}\approx\frac{\mu_1^\lambda c}{2\lambda^2}$, corresponding to coherence time $\tau_c=\frac{0.885}{\Delta f_{\text{FWHM}}}\sim3.54\tau_1$ \cite{ramelow2009discrete}. Thus, limited by the coherence time, we are able to observe interference dips at $\tau_2=\pm\tau_1$ and an interference peak at $\tau_2=0$.


\indent \emph{Dimensionality estimation of high-dimensional frequency entanglement}\rule[2pt]{8pt}{1pt}
Entanglement dimensionality is an important metric, which reveals the minimum number of dimensions that is needed to faithfully reproduce correlations of the state in any global product basis. 
Based on the observation of frequency anti-correlations and coherent superpositions in two-dimensional subspaces $\{\ket{\omega_j},\ket{\omega_{m-j}}\}$, the produced state is actually close to the form of Eq. \eqref{eq:4}.
The measurements performed serve as testbeds for our frequency-bin entangled source. For using these states in a quantum communication setting with local certification of entanglement, two further ingredients will be needed. In particular, we will need local superposition measurements of different frequency modes \cite{Lu2018,lu2020fully, Brecht2015} to enable spatially separated quantum communication protocols and we will need to ascertain the visibilities in all pairs of frequency subspaces. Nonetheless, our experiment serves as a source characterisation, showing the clear potential of the source and the notion of frequency modes for generating high-dimensional entanglement. In fact, assuming that the visibilites in all $m(m-1)/2$ frequency subspaces correspond to the average visibility of the measured two-dimensional subspaces, the achievable entanglement of formation in our experiment is $E_2=0.57$, $E_4=1.05$ and $E_6=1.56$ for the two-, four- and six-dimensional frequency modes \cite{martin2017quantifying}.

\indent \emph{Conclusion.}\rule[2pt]{8pt}{1pt}
We have presented a method of exploiting temporal distinguishability in HOM interference to generate, characterize, and verify high-dimensional entanglement in frequency-bin qudits. 
We hope that our results inspire not only practical quantum communication and quantum computation applications based on high-dimensional frequency entanglement, but also new experimental configurations exploiting quantum interference in other degrees of freedom, e.g., orbital angular momentum.

\section*{Acknowledgements}
Financial support from the Austrian Research Promotion Agency (FFG) Projects - Agentur f\"{u}r Luft- und Raumfahrt (FFG-ALR contract 6238191 and 866025), the European Space Agency (ESA contract 4000112591/14/NL/US) as well as the Austrian Academy of Sciences is gratefully acknowledged. FS acknowledges financial support from Fraunhofer Internal Programs under Grant No. Attract 066-604178. YC and LC acknowledge financial support from the National Natural Science Foundation of China (61975169), the Fundamental Research Funds for the Central Universities at Xiamen University (20720190057).
\bibliography{apssamp}

\begin{thebibliography}{30}%
\makeatletter
\providecommand \@ifxundefined [1]{%
 \@ifx{#1\undefined}
}%
\providecommand \@ifnum [1]{%
 \ifnum #1\expandafter \@firstoftwo
 \else \expandafter \@secondoftwo
 \fi
}%
\providecommand \@ifx [1]{%
 \ifx #1\expandafter \@firstoftwo
 \else \expandafter \@secondoftwo
 \fi
}%
\providecommand \natexlab [1]{#1}%
\providecommand \enquote  [1]{``#1''}%
\providecommand \bibnamefont  [1]{#1}%
\providecommand \bibfnamefont [1]{#1}%
\providecommand \citenamefont [1]{#1}%
\providecommand \href@noop [0]{\@secondoftwo}%
\providecommand \href [0]{\begingroup \@sanitize@url \@href}%
\providecommand \@href[1]{\@@startlink{#1}\@@href}%
\providecommand \@@href[1]{\endgroup#1\@@endlink}%
\providecommand \@sanitize@url [0]{\catcode `\\12\catcode `\$12\catcode
  `\&12\catcode `\#12\catcode `\^12\catcode `\_12\catcode `\%12\relax}%
\providecommand \@@startlink[1]{}%
\providecommand \@@endlink[0]{}%
\providecommand \url  [0]{\begingroup\@sanitize@url \@url }%
\providecommand \@url [1]{\endgroup\@href {#1}{\urlprefix }}%
\providecommand \urlprefix  [0]{URL }%
\providecommand \Eprint [0]{\href }%
\providecommand \doibase [0]{http://dx.doi.org/}%
\providecommand \selectlanguage [0]{\@gobble}%
\providecommand \bibinfo  [0]{\@secondoftwo}%
\providecommand \bibfield  [0]{\@secondoftwo}%
\providecommand \translation [1]{[#1]}%
\providecommand \BibitemOpen [0]{}%
\providecommand \bibitemStop [0]{}%
\providecommand \bibitemNoStop [0]{.\EOS\space}%
\providecommand \EOS [0]{\spacefactor3000\relax}%
\providecommand \BibitemShut  [1]{\csname bibitem#1\endcsname}%
\let\auto@bib@innerbib\@empty
\bibitem [{\citenamefont {Ecker}\ \emph {et~al.}(2019)\citenamefont {Ecker},
  \citenamefont {Bouchard}, \citenamefont {Bulla}, \citenamefont {Brandt},
  \citenamefont {Kohout}, \citenamefont {Steinlechner}, \citenamefont
  {Fickler}, \citenamefont {Malik}, \citenamefont {Guryanova}, \citenamefont
  {Ursin},\ and\ \citenamefont {Huber}}]{PhysRevX.9.041042}%
  \BibitemOpen
  \bibfield  {author} {\bibinfo {author} {\bibfnamefont {S.}~\bibnamefont
  {Ecker}}, \bibinfo {author} {\bibfnamefont {F.}~\bibnamefont {Bouchard}},
  \bibinfo {author} {\bibfnamefont {L.}~\bibnamefont {Bulla}}, \bibinfo
  {author} {\bibfnamefont {F.}~\bibnamefont {Brandt}}, \bibinfo {author}
  {\bibfnamefont {O.}~\bibnamefont {Kohout}}, \bibinfo {author} {\bibfnamefont
  {F.}~\bibnamefont {Steinlechner}}, \bibinfo {author} {\bibfnamefont
  {R.}~\bibnamefont {Fickler}}, \bibinfo {author} {\bibfnamefont
  {M.}~\bibnamefont {Malik}}, \bibinfo {author} {\bibfnamefont
  {Y.}~\bibnamefont {Guryanova}}, \bibinfo {author} {\bibfnamefont
  {R.}~\bibnamefont {Ursin}}, \ and\ \bibinfo {author} {\bibfnamefont
  {M.}~\bibnamefont {Huber}},\ }\href {\doibase 10.1103/PhysRevX.9.041042}
  {\bibfield  {journal} {\bibinfo  {journal} {Phys. Rev. X}\ }\textbf {\bibinfo
  {volume} {9}},\ \bibinfo {pages} {041042} (\bibinfo {year}
  {2019})}\BibitemShut {NoStop}%
\bibitem [{\citenamefont {Dada}\ \emph {et~al.}(2011)\citenamefont {Dada},
  \citenamefont {Leach}, \citenamefont {Buller}, \citenamefont {Padgett},\ and\
  \citenamefont {Andersson}}]{dada2011experimental}%
  \BibitemOpen
  \bibfield  {author} {\bibinfo {author} {\bibfnamefont {A.~C.}\ \bibnamefont
  {Dada}}, \bibinfo {author} {\bibfnamefont {J.}~\bibnamefont {Leach}},
  \bibinfo {author} {\bibfnamefont {G.~S.}\ \bibnamefont {Buller}}, \bibinfo
  {author} {\bibfnamefont {M.~J.}\ \bibnamefont {Padgett}}, \ and\ \bibinfo
  {author} {\bibfnamefont {E.}~\bibnamefont {Andersson}},\ }\href@noop {}
  {\bibfield  {journal} {\bibinfo  {journal} {Nature Physics}\ }\textbf
  {\bibinfo {volume} {7}},\ \bibinfo {pages} {677} (\bibinfo {year}
  {2011})}\BibitemShut {NoStop}%
\bibitem [{\citenamefont {Mirhosseini}\ \emph {et~al.}(2015)\citenamefont
  {Mirhosseini}, \citenamefont {Maga{\~n}a-Loaiza}, \citenamefont
  {OSullivan}, \citenamefont {Rodenburg}, \citenamefont {Malik},
  \citenamefont {Lavery}, \citenamefont {Padgett}, \citenamefont {Gauthier},\
  and\ \citenamefont {Boyd}}]{mirhosseini2015high}%
  \BibitemOpen
  \bibfield  {author} {\bibinfo {author} {\bibfnamefont {M.}~\bibnamefont
  {Mirhosseini}}, \bibinfo {author} {\bibfnamefont {O.~S.}\ \bibnamefont
  {Maga{\~n}a-Loaiza}}, \bibinfo {author} {\bibfnamefont {M.~N.}\ \bibnamefont
  {OSullivan}}, \bibinfo {author} {\bibfnamefont {B.}~\bibnamefont
  {Rodenburg}}, \bibinfo {author} {\bibfnamefont {M.}~\bibnamefont {Malik}},
  \bibinfo {author} {\bibfnamefont {M.~P.}\ \bibnamefont {Lavery}}, \bibinfo
  {author} {\bibfnamefont {M.~J.}\ \bibnamefont {Padgett}}, \bibinfo {author}
  {\bibfnamefont {D.~J.}\ \bibnamefont {Gauthier}}, \ and\ \bibinfo {author}
  {\bibfnamefont {R.~W.}\ \bibnamefont {Boyd}},\ }\href@noop {} {\bibfield
  {journal} {\bibinfo  {journal} {New Journal of Physics}\ }\textbf {\bibinfo
  {volume} {17}},\ \bibinfo {pages} {033033} (\bibinfo {year}
  {2015})}\BibitemShut {NoStop}%
\bibitem [{\citenamefont {Raussendorf}\ and\ \citenamefont
  {Harrington}(2007)}]{raussendorf2007fault}%
  \BibitemOpen
  \bibfield  {author} {\bibinfo {author} {\bibfnamefont {R.}~\bibnamefont
  {Raussendorf}}\ and\ \bibinfo {author} {\bibfnamefont {J.}~\bibnamefont
  {Harrington}},\ }\href@noop {} {\bibfield  {journal} {\bibinfo  {journal}
  {Physical review letters}\ }\textbf {\bibinfo {volume} {98}},\ \bibinfo
  {pages} {190504} (\bibinfo {year} {2007})}\BibitemShut {NoStop}%
\bibitem [{\citenamefont {Molina-Terriza}\ \emph {et~al.}(2007)\citenamefont
  {Molina-Terriza}, \citenamefont {Torres},\ and\ \citenamefont
  {Torner}}]{molina2007twisted}%
  \BibitemOpen
  \bibfield  {author} {\bibinfo {author} {\bibfnamefont {G.}~\bibnamefont
  {Molina-Terriza}}, \bibinfo {author} {\bibfnamefont {J.~P.}\ \bibnamefont
  {Torres}}, \ and\ \bibinfo {author} {\bibfnamefont {L.}~\bibnamefont
  {Torner}},\ }\href@noop {} {\bibfield  {journal} {\bibinfo  {journal} {Nature
  physics}\ }\textbf {\bibinfo {volume} {3}},\ \bibinfo {pages} {305} (\bibinfo
  {year} {2007})}\BibitemShut {NoStop}%
\bibitem [{\citenamefont {Zhong}\ \emph {et~al.}(2015)\citenamefont {Zhong},
  \citenamefont {Zhou}, \citenamefont {Horansky}, \citenamefont {Lee},
  \citenamefont {Verma}, \citenamefont {Lita}, \citenamefont {Restelli},
  \citenamefont {Bienfang}, \citenamefont {Mirin}, \citenamefont {Gerrits}
  \emph {et~al.}}]{zhong2015photon}%
  \BibitemOpen
  \bibfield  {author} {\bibinfo {author} {\bibfnamefont {T.}~\bibnamefont
  {Zhong}}, \bibinfo {author} {\bibfnamefont {H.}~\bibnamefont {Zhou}},
  \bibinfo {author} {\bibfnamefont {R.~D.}\ \bibnamefont {Horansky}}, \bibinfo
  {author} {\bibfnamefont {C.}~\bibnamefont {Lee}}, \bibinfo {author}
  {\bibfnamefont {V.~B.}\ \bibnamefont {Verma}}, \bibinfo {author}
  {\bibfnamefont {A.~E.}\ \bibnamefont {Lita}}, \bibinfo {author}
  {\bibfnamefont {A.}~\bibnamefont {Restelli}}, \bibinfo {author}
  {\bibfnamefont {J.~C.}\ \bibnamefont {Bienfang}}, \bibinfo {author}
  {\bibfnamefont {R.~P.}\ \bibnamefont {Mirin}}, \bibinfo {author}
  {\bibfnamefont {T.}~\bibnamefont {Gerrits}},  \emph {et~al.},\ }\href@noop {}
  {\bibfield  {journal} {\bibinfo  {journal} {New Journal of Physics}\ }\textbf
  {\bibinfo {volume} {17}},\ \bibinfo {pages} {022002} (\bibinfo {year}
  {2015})}\BibitemShut {NoStop}%
\bibitem [{\citenamefont {Fickler}\ \emph {et~al.}(2014)\citenamefont
  {Fickler}, \citenamefont {Lapkiewicz}, \citenamefont {Huber}, \citenamefont
  {Lavery}, \citenamefont {Padgett},\ and\ \citenamefont
  {Zeilinger}}]{fickler2014interface}%
  \BibitemOpen
  \bibfield  {author} {\bibinfo {author} {\bibfnamefont {R.}~\bibnamefont
  {Fickler}}, \bibinfo {author} {\bibfnamefont {R.}~\bibnamefont {Lapkiewicz}},
  \bibinfo {author} {\bibfnamefont {M.}~\bibnamefont {Huber}}, \bibinfo
  {author} {\bibfnamefont {M.~P.}\ \bibnamefont {Lavery}}, \bibinfo {author}
  {\bibfnamefont {M.~J.}\ \bibnamefont {Padgett}}, \ and\ \bibinfo {author}
  {\bibfnamefont {A.}~\bibnamefont {Zeilinger}},\ }\href@noop {} {\bibfield
  {journal} {\bibinfo  {journal} {Nature communications}\ }\textbf {\bibinfo
  {volume} {5}},\ \bibinfo {pages} {4502} (\bibinfo {year} {2014})}\BibitemShut
  {NoStop}%
\bibitem [{\citenamefont {Kovlakov}\ \emph {et~al.}(2017)\citenamefont
  {Kovlakov}, \citenamefont {Bobrov}, \citenamefont {Straupe},\ and\
  \citenamefont {Kulik}}]{kovlakov2017spatial}%
  \BibitemOpen
  \bibfield  {author} {\bibinfo {author} {\bibfnamefont {E.}~\bibnamefont
  {Kovlakov}}, \bibinfo {author} {\bibfnamefont {I.}~\bibnamefont {Bobrov}},
  \bibinfo {author} {\bibfnamefont {S.}~\bibnamefont {Straupe}}, \ and\
  \bibinfo {author} {\bibfnamefont {S.}~\bibnamefont {Kulik}},\ }\href@noop {}
  {\bibfield  {journal} {\bibinfo  {journal} {Physical review letters}\
  }\textbf {\bibinfo {volume} {118}},\ \bibinfo {pages} {030503} (\bibinfo
  {year} {2017})}\BibitemShut {NoStop}%
\bibitem [{\citenamefont {Steinlechner}\ \emph {et~al.}(2017)\citenamefont
  {Steinlechner}, \citenamefont {Ecker}, \citenamefont {Fink}, \citenamefont
  {Liu}, \citenamefont {Bavaresco}, \citenamefont {Huber}, \citenamefont
  {Scheidl},\ and\ \citenamefont {Ursin}}]{Steinlechner2017}%
  \BibitemOpen
  \bibfield  {author} {\bibinfo {author} {\bibfnamefont {F.}~\bibnamefont
  {Steinlechner}}, \bibinfo {author} {\bibfnamefont {S.}~\bibnamefont {Ecker}},
  \bibinfo {author} {\bibfnamefont {M.}~\bibnamefont {Fink}}, \bibinfo {author}
  {\bibfnamefont {B.}~\bibnamefont {Liu}}, \bibinfo {author} {\bibfnamefont
  {J.}~\bibnamefont {Bavaresco}}, \bibinfo {author} {\bibfnamefont
  {M.}~\bibnamefont {Huber}}, \bibinfo {author} {\bibfnamefont
  {T.}~\bibnamefont {Scheidl}}, \ and\ \bibinfo {author} {\bibfnamefont
  {R.}~\bibnamefont {Ursin}},\ }\href {\doibase 10.1038/ncomms15971} {\bibfield
   {journal} {\bibinfo  {journal} {Nature Communications}\ }\textbf {\bibinfo
  {volume} {8}},\ \bibinfo {pages} {15971} (\bibinfo {year}
  {2017})}\BibitemShut {NoStop}%
\bibitem [{\citenamefont {Kuzucu}\ \emph {et~al.}(2005)\citenamefont {Kuzucu},
  \citenamefont {Fiorentino}, \citenamefont {Albota}, \citenamefont {Wong},\
  and\ \citenamefont {K{\"a}rtner}}]{kuzucu2005two}%
  \BibitemOpen
  \bibfield  {author} {\bibinfo {author} {\bibfnamefont {O.}~\bibnamefont
  {Kuzucu}}, \bibinfo {author} {\bibfnamefont {M.}~\bibnamefont {Fiorentino}},
  \bibinfo {author} {\bibfnamefont {M.~A.}\ \bibnamefont {Albota}}, \bibinfo
  {author} {\bibfnamefont {F.~N.}\ \bibnamefont {Wong}}, \ and\ \bibinfo
  {author} {\bibfnamefont {F.~X.}\ \bibnamefont {K{\"a}rtner}},\ }\href@noop {}
  {\bibfield  {journal} {\bibinfo  {journal} {Physical review letters}\
  }\textbf {\bibinfo {volume} {94}},\ \bibinfo {pages} {083601} (\bibinfo
  {year} {2005})}\BibitemShut {NoStop}%
\bibitem [{\citenamefont {Roslund}\ \emph {et~al.}(2014)\citenamefont
  {Roslund}, \citenamefont {De~Araujo}, \citenamefont {Jiang}, \citenamefont
  {Fabre},\ and\ \citenamefont {Treps}}]{roslund2014wavelength}%
  \BibitemOpen
  \bibfield  {author} {\bibinfo {author} {\bibfnamefont {J.}~\bibnamefont
  {Roslund}}, \bibinfo {author} {\bibfnamefont {R.~M.}\ \bibnamefont
  {De~Araujo}}, \bibinfo {author} {\bibfnamefont {S.}~\bibnamefont {Jiang}},
  \bibinfo {author} {\bibfnamefont {C.}~\bibnamefont {Fabre}}, \ and\ \bibinfo
  {author} {\bibfnamefont {N.}~\bibnamefont {Treps}},\ }\href@noop {}
  {\bibfield  {journal} {\bibinfo  {journal} {Nature Photonics}\ }\textbf
  {\bibinfo {volume} {8}},\ \bibinfo {pages} {109} (\bibinfo {year}
  {2014})}\BibitemShut {NoStop}%
\bibitem [{\citenamefont {Yan}\ \emph {et~al.}(2011)\citenamefont {Yan},
  \citenamefont {Zhang}, \citenamefont {Chen}, \citenamefont {Loy},
  \citenamefont {Wong},\ and\ \citenamefont {Du}}]{yan2011generation}%
  \BibitemOpen
  \bibfield  {author} {\bibinfo {author} {\bibfnamefont {H.}~\bibnamefont
  {Yan}}, \bibinfo {author} {\bibfnamefont {S.}~\bibnamefont {Zhang}}, \bibinfo
  {author} {\bibfnamefont {J.}~\bibnamefont {Chen}}, \bibinfo {author}
  {\bibfnamefont {M.~M.}\ \bibnamefont {Loy}}, \bibinfo {author} {\bibfnamefont
  {G.~K.}\ \bibnamefont {Wong}}, \ and\ \bibinfo {author} {\bibfnamefont
  {S.}~\bibnamefont {Du}},\ }\href@noop {} {\bibfield  {journal} {\bibinfo
  {journal} {Physical review letters}\ }\textbf {\bibinfo {volume} {106}},\
  \bibinfo {pages} {033601} (\bibinfo {year} {2011})}\BibitemShut {NoStop}%
\bibitem [{\citenamefont {{Lingaraju}}\ \emph {et~al.}(2019)\citenamefont
  {{Lingaraju}}, \citenamefont {{Lu}}, \citenamefont {{Seshadri}},
  \citenamefont {{Imany}}, \citenamefont {{Leaird}}, \citenamefont {{Lukens}},\
  and\ \citenamefont {{Weiner}}}]{lingaraju2019quantum}%
  \BibitemOpen
  \bibfield  {author} {\bibinfo {author} {\bibfnamefont {N.~B.}\ \bibnamefont
  {{Lingaraju}}}, \bibinfo {author} {\bibfnamefont {H.-H.}\ \bibnamefont
  {{Lu}}}, \bibinfo {author} {\bibfnamefont {S.}~\bibnamefont {{Seshadri}}},
  \bibinfo {author} {\bibfnamefont {P.}~\bibnamefont {{Imany}}}, \bibinfo
  {author} {\bibfnamefont {D.~E.}\ \bibnamefont {{Leaird}}}, \bibinfo {author}
  {\bibfnamefont {J.~M.}\ \bibnamefont {{Lukens}}}, \ and\ \bibinfo {author}
  {\bibfnamefont {A.~M.}\ \bibnamefont {{Weiner}}},\ }\href@noop {} {\bibfield
  {journal} {\bibinfo  {journal} {Optics Express}\ }\textbf {\bibinfo {volume}
  {27}},\ \bibinfo {pages} {38683} (\bibinfo {year} {2019})}\BibitemShut
  {NoStop}%
\bibitem [{\citenamefont {Orre}\ \emph {et~al.}(2019)\citenamefont {Orre},
  \citenamefont {Goldschmidt}, \citenamefont {Deshpande}, \citenamefont
  {Gorshkov}, \citenamefont {Tamma}, \citenamefont {Hafezi},\ and\
  \citenamefont {Mittal}}]{venkata2019interfernce}%
  \BibitemOpen
  \bibfield  {author} {\bibinfo {author} {\bibfnamefont {V.~V.}\ \bibnamefont
  {Orre}}, \bibinfo {author} {\bibfnamefont {E.~A.}\ \bibnamefont
  {Goldschmidt}}, \bibinfo {author} {\bibfnamefont {A.}~\bibnamefont
  {Deshpande}}, \bibinfo {author} {\bibfnamefont {A.~V.}\ \bibnamefont
  {Gorshkov}}, \bibinfo {author} {\bibfnamefont {V.}~\bibnamefont {Tamma}},
  \bibinfo {author} {\bibfnamefont {M.}~\bibnamefont {Hafezi}}, \ and\ \bibinfo
  {author} {\bibfnamefont {S.}~\bibnamefont {Mittal}},\ }\href {\doibase
  10.1103/PhysRevLett.123.123603} {\bibfield  {journal} {\bibinfo  {journal}
  {Phys. Rev. Lett.}\ }\textbf {\bibinfo {volume} {123}},\ \bibinfo {pages}
  {123603} (\bibinfo {year} {2019})}\BibitemShut {NoStop}%
\bibitem [{\citenamefont {Sören}\ \emph {et~al.}(2018)\citenamefont {Sören},
  \citenamefont {Wengerowsky}, \citenamefont {Siddarth}, \citenamefont
  {Koduru}, \citenamefont {Joshi}, \citenamefont {Fabian}, \citenamefont
  {Steinlechner}, \citenamefont {Hannes}, \citenamefont {Hübel},\ and\
  \citenamefont {and}}]{soren2018entanglement}%
  \BibitemOpen
  \bibfield  {author} {\bibinfo {author} {\bibnamefont {Sören}}, \bibinfo
  {author} {\bibnamefont {Wengerowsky}}, \bibinfo {author} {\bibnamefont
  {Siddarth}}, \bibinfo {author} {\bibnamefont {Koduru}}, \bibinfo {author}
  {\bibnamefont {Joshi}}, \bibinfo {author} {\bibnamefont {Fabian}}, \bibinfo
  {author} {\bibnamefont {Steinlechner}}, \bibinfo {author} {\bibnamefont
  {Hannes}}, \bibinfo {author} {\bibnamefont {Hübel}}, \ and\ \bibinfo
  {author} {\bibfnamefont {R.}~\bibnamefont {and}},\ }\href {\doibase
  10.1038/s41586-018-0766-y} {\bibfield  {journal} {\bibinfo  {journal}
  {Nature}\ } (\bibinfo {year} {2018}),\ 10.1038/s41586-018-0766-y}\BibitemShut
  {NoStop}%
\bibitem [{\citenamefont {Steinlechner}\ \emph {et~al.}(2012)\citenamefont
  {Steinlechner}, \citenamefont {Trojek}, \citenamefont {Jofre}, \citenamefont
  {Weier}, \citenamefont {Perez}, \citenamefont {Jennewein}, \citenamefont
  {Ursin}, \citenamefont {Rarity}, \citenamefont {Mitchell}, \citenamefont
  {Torres}, \citenamefont {Weinfurter},\ and\ \citenamefont
  {Pruneri}}]{fabian2012high}%
  \BibitemOpen
  \bibfield  {author} {\bibinfo {author} {\bibfnamefont {F.}~\bibnamefont
  {Steinlechner}}, \bibinfo {author} {\bibfnamefont {P.}~\bibnamefont
  {Trojek}}, \bibinfo {author} {\bibfnamefont {M.}~\bibnamefont {Jofre}},
  \bibinfo {author} {\bibfnamefont {H.}~\bibnamefont {Weier}}, \bibinfo
  {author} {\bibfnamefont {D.}~\bibnamefont {Perez}}, \bibinfo {author}
  {\bibfnamefont {T.}~\bibnamefont {Jennewein}}, \bibinfo {author}
  {\bibfnamefont {R.}~\bibnamefont {Ursin}}, \bibinfo {author} {\bibfnamefont
  {J.}~\bibnamefont {Rarity}}, \bibinfo {author} {\bibfnamefont {M.~W.}\
  \bibnamefont {Mitchell}}, \bibinfo {author} {\bibfnamefont {J.~P.}\
  \bibnamefont {Torres}}, \bibinfo {author} {\bibfnamefont {H.}~\bibnamefont
  {Weinfurter}}, \ and\ \bibinfo {author} {\bibfnamefont {V.}~\bibnamefont
  {Pruneri}},\ }\href {\doibase 10.1364/OE.20.009640} {\bibfield  {journal}
  {\bibinfo  {journal} {Opt. Express}\ }\textbf {\bibinfo {volume} {20}},\
  \bibinfo {pages} {9640} (\bibinfo {year} {2012})}\BibitemShut {NoStop}%
\bibitem [{\citenamefont {Fedrizzi}\ \emph {et~al.}(2007)\citenamefont
  {Fedrizzi}, \citenamefont {Herbst}, \citenamefont {Poppe}, \citenamefont
  {Jennewein},\ and\ \citenamefont {Zeilinger}}]{fedrizzi2007wavelength}%
  \BibitemOpen
  \bibfield  {author} {\bibinfo {author} {\bibfnamefont {A.}~\bibnamefont
  {Fedrizzi}}, \bibinfo {author} {\bibfnamefont {T.}~\bibnamefont {Herbst}},
  \bibinfo {author} {\bibfnamefont {A.}~\bibnamefont {Poppe}}, \bibinfo
  {author} {\bibfnamefont {T.}~\bibnamefont {Jennewein}}, \ and\ \bibinfo
  {author} {\bibfnamefont {A.}~\bibnamefont {Zeilinger}},\ }\href {\doibase
  10.1364/OE.15.015377} {\bibfield  {journal} {\bibinfo  {journal} {Opt.
  Express}\ }\textbf {\bibinfo {volume} {15}},\ \bibinfo {pages} {15377}
  (\bibinfo {year} {2007})}\BibitemShut {NoStop}%
\bibitem [{\citenamefont {Kwiat}\ \emph {et~al.}(1999)\citenamefont {Kwiat},
  \citenamefont {Waks}, \citenamefont {White}, \citenamefont {Appelbaum},\ and\
  \citenamefont {Eberhard}}]{kwiat1999ultrabright}%
  \BibitemOpen
  \bibfield  {author} {\bibinfo {author} {\bibfnamefont {P.~G.}\ \bibnamefont
  {Kwiat}}, \bibinfo {author} {\bibfnamefont {E.}~\bibnamefont {Waks}},
  \bibinfo {author} {\bibfnamefont {A.~G.}\ \bibnamefont {White}}, \bibinfo
  {author} {\bibfnamefont {I.}~\bibnamefont {Appelbaum}}, \ and\ \bibinfo
  {author} {\bibfnamefont {P.~H.}\ \bibnamefont {Eberhard}},\ }\href {\doibase
  10.1103/PhysRevA.60.R773} {\bibfield  {journal} {\bibinfo  {journal} {Phys.
  Rev. A}\ }\textbf {\bibinfo {volume} {60}},\ \bibinfo {pages} {R773}
  (\bibinfo {year} {1999})}\BibitemShut {NoStop}%
\bibitem [{\citenamefont {Chen}\ \emph {et~al.}(2018)\citenamefont {Chen},
  \citenamefont {Ecker}, \citenamefont {Wengerowsky}, \citenamefont {Bulla},
  \citenamefont {Joshi}, \citenamefont {Steinlechner},\ and\ \citenamefont
  {Ursin}}]{chen2018polarization}%
  \BibitemOpen
  \bibfield  {author} {\bibinfo {author} {\bibfnamefont {Y.}~\bibnamefont
  {Chen}}, \bibinfo {author} {\bibfnamefont {S.}~\bibnamefont {Ecker}},
  \bibinfo {author} {\bibfnamefont {S.}~\bibnamefont {Wengerowsky}}, \bibinfo
  {author} {\bibfnamefont {L.}~\bibnamefont {Bulla}}, \bibinfo {author}
  {\bibfnamefont {S.~K.}\ \bibnamefont {Joshi}}, \bibinfo {author}
  {\bibfnamefont {F.}~\bibnamefont {Steinlechner}}, \ and\ \bibinfo {author}
  {\bibfnamefont {R.}~\bibnamefont {Ursin}},\ }\href {\doibase
  10.1103/PhysRevLett.121.200502} {\bibfield  {journal} {\bibinfo  {journal}
  {Phys. Rev. Lett.}\ }\textbf {\bibinfo {volume} {121}},\ \bibinfo {pages}
  {200502} (\bibinfo {year} {2018})}\BibitemShut {NoStop}%
\bibitem [{\citenamefont {Jin}\ \emph {et~al.}(2016)\citenamefont {Jin},
  \citenamefont {Shimizu}, \citenamefont {Fujiwara}, \citenamefont {Takeoka},
  \citenamefont {Wakabayashi}, \citenamefont {Yamashita}, \citenamefont {Miki},
  \citenamefont {Terai}, \citenamefont {Gerrits},\ and\ \citenamefont
  {Sasaki}}]{jin2016simple}%
  \BibitemOpen
  \bibfield  {author} {\bibinfo {author} {\bibfnamefont {R.-B.}\ \bibnamefont
  {Jin}}, \bibinfo {author} {\bibfnamefont {R.}~\bibnamefont {Shimizu}},
  \bibinfo {author} {\bibfnamefont {M.}~\bibnamefont {Fujiwara}}, \bibinfo
  {author} {\bibfnamefont {M.}~\bibnamefont {Takeoka}}, \bibinfo {author}
  {\bibfnamefont {R.}~\bibnamefont {Wakabayashi}}, \bibinfo {author}
  {\bibfnamefont {T.}~\bibnamefont {Yamashita}}, \bibinfo {author}
  {\bibfnamefont {S.}~\bibnamefont {Miki}}, \bibinfo {author} {\bibfnamefont
  {H.}~\bibnamefont {Terai}}, \bibinfo {author} {\bibfnamefont
  {T.}~\bibnamefont {Gerrits}}, \ and\ \bibinfo {author} {\bibfnamefont
  {M.}~\bibnamefont {Sasaki}},\ }\href@noop {} {\bibfield  {journal} {\bibinfo
  {journal} {Quantum Science and Technology}\ }\textbf {\bibinfo {volume}
  {1}},\ \bibinfo {pages} {015004} (\bibinfo {year} {2016})}\BibitemShut
  {NoStop}%
\bibitem [{\citenamefont {Hong}\ \emph {et~al.}(1987)\citenamefont {Hong},
  \citenamefont {Ou},\ and\ \citenamefont {Mandel}}]{hong1987measurement}%
  \BibitemOpen
  \bibfield  {author} {\bibinfo {author} {\bibfnamefont {C.~K.}\ \bibnamefont
  {Hong}}, \bibinfo {author} {\bibfnamefont {Z.~Y.}\ \bibnamefont {Ou}}, \ and\
  \bibinfo {author} {\bibfnamefont {L.}~\bibnamefont {Mandel}},\ }\href
  {\doibase 10.1103/PhysRevLett.59.2044} {\bibfield  {journal} {\bibinfo
  {journal} {Phys. Rev. Lett.}\ }\textbf {\bibinfo {volume} {59}},\ \bibinfo
  {pages} {2044} (\bibinfo {year} {1987})}\BibitemShut {NoStop}%
\bibitem [{\citenamefont {Kues}\ \emph {et~al.}(2017)\citenamefont {Kues},
  \citenamefont {Reimer}, \citenamefont {Roztocki}, \citenamefont {Cort{\'e}s},
  \citenamefont {Sciara}, \citenamefont {Wetzel}, \citenamefont {Zhang},
  \citenamefont {Cino}, \citenamefont {Chu}, \citenamefont {Little} \emph
  {et~al.}}]{kues2017chip}%
  \BibitemOpen
  \bibfield  {author} {\bibinfo {author} {\bibfnamefont {M.}~\bibnamefont
  {Kues}}, \bibinfo {author} {\bibfnamefont {C.}~\bibnamefont {Reimer}},
  \bibinfo {author} {\bibfnamefont {P.}~\bibnamefont {Roztocki}}, \bibinfo
  {author} {\bibfnamefont {L.~R.}\ \bibnamefont {Cort{\'e}s}}, \bibinfo
  {author} {\bibfnamefont {S.}~\bibnamefont {Sciara}}, \bibinfo {author}
  {\bibfnamefont {B.}~\bibnamefont {Wetzel}}, \bibinfo {author} {\bibfnamefont
  {Y.}~\bibnamefont {Zhang}}, \bibinfo {author} {\bibfnamefont
  {A.}~\bibnamefont {Cino}}, \bibinfo {author} {\bibfnamefont {S.~T.}\
  \bibnamefont {Chu}}, \bibinfo {author} {\bibfnamefont {B.~E.}\ \bibnamefont
  {Little}},  \emph {et~al.},\ }\href@noop {} {\bibfield  {journal} {\bibinfo
  {journal} {Nature}\ }\textbf {\bibinfo {volume} {546}},\ \bibinfo {pages}
  {622} (\bibinfo {year} {2017})}\BibitemShut {NoStop}%
\bibitem [{\citenamefont {Lu}\ \emph {et~al.}(2018)\citenamefont {Lu},
  \citenamefont {Lukens}, \citenamefont {Peters}, \citenamefont {Odele},
  \citenamefont {Leaird}, \citenamefont {Weiner},\ and\ \citenamefont
  {Lougovski}}]{Lu2018}%
  \BibitemOpen
  \bibfield  {author} {\bibinfo {author} {\bibfnamefont {H.-H.}\ \bibnamefont
  {Lu}}, \bibinfo {author} {\bibfnamefont {J.~M.}\ \bibnamefont {Lukens}},
  \bibinfo {author} {\bibfnamefont {N.~A.}\ \bibnamefont {Peters}}, \bibinfo
  {author} {\bibfnamefont {O.~D.}\ \bibnamefont {Odele}}, \bibinfo {author}
  {\bibfnamefont {D.~E.}\ \bibnamefont {Leaird}}, \bibinfo {author}
  {\bibfnamefont {A.~M.}\ \bibnamefont {Weiner}}, \ and\ \bibinfo {author}
  {\bibfnamefont {P.}~\bibnamefont {Lougovski}},\ }\href {\doibase
  10.1103/PhysRevLett.120.030502} {\bibfield  {journal} {\bibinfo  {journal}
  {Physical Review Letters}\ }\textbf {\bibinfo {volume} {120}},\ \bibinfo
  {pages} {030502} (\bibinfo {year} {2018})}\BibitemShut {NoStop}%
\bibitem [{\citenamefont {Maclean}\ \emph {et~al.}(2018)\citenamefont
  {Maclean}, \citenamefont {Donohue},\ and\ \citenamefont
  {Resch}}]{Maclean2018}%
  \BibitemOpen
  \bibfield  {author} {\bibinfo {author} {\bibfnamefont {J.~P.~W.}\
  \bibnamefont {Maclean}}, \bibinfo {author} {\bibfnamefont {J.~M.}\
  \bibnamefont {Donohue}}, \ and\ \bibinfo {author} {\bibfnamefont {K.~J.}\
  \bibnamefont {Resch}},\ }\href {\doibase 10.1103/PhysRevLett.120.053601}
  {\bibfield  {journal} {\bibinfo  {journal} {Physical Review Letters}\
  }\textbf {\bibinfo {volume} {120}},\ \bibinfo {pages} {53601} (\bibinfo
  {year} {2018})}\BibitemShut {NoStop}%
\bibitem [{\citenamefont {Brecht}\ \emph {et~al.}(2015)\citenamefont {Brecht},
  \citenamefont {Reddy}, \citenamefont {Silberhorn},\ and\ \citenamefont
  {Raymer}}]{Brecht2015}%
  \BibitemOpen
  \bibfield  {author} {\bibinfo {author} {\bibfnamefont {B.}~\bibnamefont
  {Brecht}}, \bibinfo {author} {\bibfnamefont {D.~V.}\ \bibnamefont {Reddy}},
  \bibinfo {author} {\bibfnamefont {C.}~\bibnamefont {Silberhorn}}, \ and\
  \bibinfo {author} {\bibfnamefont {M.~G.}\ \bibnamefont {Raymer}},\ }\href
  {\doibase 10.1103/PhysRevX.5.041017} {\bibfield  {journal} {\bibinfo
  {journal} {Physical Review X}\ }\textbf {\bibinfo {volume} {5}},\ \bibinfo
  {pages} {1} (\bibinfo {year} {2015})}\BibitemShut {NoStop}%
\bibitem [{\citenamefont {Ou}\ and\ \citenamefont
  {Mandel}(1988)}]{ou1988observation}%
  \BibitemOpen
  \bibfield  {author} {\bibinfo {author} {\bibfnamefont {Z.~Y.}\ \bibnamefont
  {Ou}}\ and\ \bibinfo {author} {\bibfnamefont {L.}~\bibnamefont {Mandel}},\
  }\href {\doibase 10.1103/PhysRevLett.61.54} {\bibfield  {journal} {\bibinfo
  {journal} {Phys. Rev. Lett.}\ }\textbf {\bibinfo {volume} {61}},\ \bibinfo
  {pages} {54} (\bibinfo {year} {1988})}\BibitemShut {NoStop}%
\bibitem [{\citenamefont {Fedrizzi}\ \emph {et~al.}(2009)\citenamefont
  {Fedrizzi}, \citenamefont {Herbst}, \citenamefont {Aspelmeyer}, \citenamefont
  {Barbieri}, \citenamefont {Jennewein},\ and\ \citenamefont
  {Zeilinger}}]{fedrizzi2009anti}%
  \BibitemOpen
  \bibfield  {author} {\bibinfo {author} {\bibfnamefont {A.}~\bibnamefont
  {Fedrizzi}}, \bibinfo {author} {\bibfnamefont {T.}~\bibnamefont {Herbst}},
  \bibinfo {author} {\bibfnamefont {M.}~\bibnamefont {Aspelmeyer}}, \bibinfo
  {author} {\bibfnamefont {M.}~\bibnamefont {Barbieri}}, \bibinfo {author}
  {\bibfnamefont {T.}~\bibnamefont {Jennewein}}, \ and\ \bibinfo {author}
  {\bibfnamefont {A.}~\bibnamefont {Zeilinger}},\ }\href@noop {} {\bibfield
  {journal} {\bibinfo  {journal} {New Journal of Physics}\ }\textbf {\bibinfo
  {volume} {11}},\ \bibinfo {pages} {103052} (\bibinfo {year}
  {2009})}\BibitemShut {NoStop}%
\bibitem [{\citenamefont {Ramelow}\ \emph {et~al.}(2009)\citenamefont
  {Ramelow}, \citenamefont {Ratschbacher}, \citenamefont {Fedrizzi},
  \citenamefont {Langford},\ and\ \citenamefont
  {Zeilinger}}]{ramelow2009discrete}%
  \BibitemOpen
  \bibfield  {author} {\bibinfo {author} {\bibfnamefont {S.}~\bibnamefont
  {Ramelow}}, \bibinfo {author} {\bibfnamefont {L.}~\bibnamefont
  {Ratschbacher}}, \bibinfo {author} {\bibfnamefont {A.}~\bibnamefont
  {Fedrizzi}}, \bibinfo {author} {\bibfnamefont {N.}~\bibnamefont {Langford}},
  \ and\ \bibinfo {author} {\bibfnamefont {A.}~\bibnamefont {Zeilinger}},\
  }\href@noop {} {\bibfield  {journal} {\bibinfo  {journal} {Physical review
  letters}\ }\textbf {\bibinfo {volume} {103}},\ \bibinfo {pages} {253601}
  (\bibinfo {year} {2009})}\BibitemShut {NoStop}%
\bibitem [{\citenamefont {Lu}\ \emph {et~al.}(2020)\citenamefont {Lu},
  \citenamefont {Simmerman}, \citenamefont {Lougovski}, \citenamefont
  {Weiner},\ and\ \citenamefont {Lukens}}]{lu2020fully}%
  \BibitemOpen
  \bibfield  {author} {\bibinfo {author} {\bibfnamefont {H.-H.}\ \bibnamefont
  {Lu}}, \bibinfo {author} {\bibfnamefont {E.~M.}\ \bibnamefont {Simmerman}},
  \bibinfo {author} {\bibfnamefont {P.}~\bibnamefont {Lougovski}}, \bibinfo
  {author} {\bibfnamefont {A.~M.}\ \bibnamefont {Weiner}}, \ and\ \bibinfo
  {author} {\bibfnamefont {J.~M.}\ \bibnamefont {Lukens}},\ }\href@noop {}
  {\enquote {\bibinfo {title} {Fully arbitrary control of frequency-bin
  qubits},}\ } (\bibinfo {year} {2020}),\ \Eprint
  {http://arxiv.org/abs/2008.07444} {arXiv:2008.07444 [quant-ph]} \BibitemShut
  {NoStop}%
\bibitem [{\citenamefont {Martin}\ \emph {et~al.}(2017)\citenamefont {Martin},
  \citenamefont {Guerreiro}, \citenamefont {Tiranov}, \citenamefont
  {Designolle}, \citenamefont {Fr{\"o}wis}, \citenamefont {Brunner},
  \citenamefont {Huber},\ and\ \citenamefont {Gisin}}]{martin2017quantifying}%
  \BibitemOpen
  \bibfield  {author} {\bibinfo {author} {\bibfnamefont {A.}~\bibnamefont
  {Martin}}, \bibinfo {author} {\bibfnamefont {T.}~\bibnamefont {Guerreiro}},
  \bibinfo {author} {\bibfnamefont {A.}~\bibnamefont {Tiranov}}, \bibinfo
  {author} {\bibfnamefont {S.}~\bibnamefont {Designolle}}, \bibinfo {author}
  {\bibfnamefont {F.}~\bibnamefont {Fr{\"o}wis}}, \bibinfo {author}
  {\bibfnamefont {N.}~\bibnamefont {Brunner}}, \bibinfo {author} {\bibfnamefont
  {M.}~\bibnamefont {Huber}}, \ and\ \bibinfo {author} {\bibfnamefont
  {N.}~\bibnamefont {Gisin}},\ }\href@noop {} {\bibfield  {journal} {\bibinfo
  {journal} {Physical review letters}\ }\textbf {\bibinfo {volume} {118}},\
  \bibinfo {pages} {110501} (\bibinfo {year} {2017})}\BibitemShut {NoStop}%
\end{thebibliography}%
\end{document}